\documentclass[a4paper,11pt]{article}
\usepackage[left=20mm,right=20mm,top=20mm,bottom=20mm]{geometry}
\usepackage{amsmath}
\usepackage{amsfonts}
\usepackage{amssymb}
\usepackage{amsbsy}
\usepackage{color,graphics}
\usepackage{setspace}
\usepackage{epsfig}
\usepackage{subfigure}
\usepackage{cite}
\usepackage{graphicx}
\usepackage{latexsym,multicol}
\usepackage{tcolorbox}

\graphicspath{{./LCE-stripes-figs/}}
\begin{document}
\title{Likely striping in stochastic nematic elastomers}
\author{L. Angela Mihai\footnote{School of Mathematics, Cardiff University, Senghennydd Road, Cardiff, CF24 4AG, UK, Email: \texttt{MihaiLA@cardiff.ac.uk}}
\qquad Alain Goriely\footnote{Mathematical Institute, University of Oxford, Woodstock Road, Oxford, OX2 6GG, UK, Email: \texttt{goriely@maths.ox.ac.uk}}
}	
\maketitle

\begin{abstract}
For monodomain nematic elastomers, we construct generalised elastic-nematic constitutive models combining purely elastic and neoclassical-type strain-energy densities. Inspired by recent developments in stochastic elasticity, we extend these models to stochastic-elastic-nematic forms where the model parameters are defined by spatially-independent probability density functions at a continuum level. To investigate the  behaviour of these systems and demonstrate the effects of the probabilistic parameters, we focus on the classical problem of shear striping in a stretched nematic elastomer for which the solution is given explicitly. We find that, unlike in the neoclassical case where the inhomogeneous deformation occurs within a universal interval that is independent of the elastic modulus, for the elastic-nematic models, the critical interval depends on the material parameters. For the stochastic extension, the bounds of this interval are probabilistic, and the homogeneous and inhomogeneous states compete in the sense that both have a a given probability to occur. We refer to the inhomogeneous pattern within this interval as `likely striping'.\\

\end{abstract}


\begin{quote}
	``Instead of stating the positions and velocities of all the molecules, we allow the possibility that these may vary for some reason - be it because we lack precise information, be it because we wish only some average in time or in space, be it because we are content to represent the result of averaging over many repetitions [...] We can then assign a probability to each quantity and calculate the values expected according to that probability.'' - C. Truesdell \cite[p.~74]{Truesdell:1984}
\end{quote}

\section{Introduction}

Nematic elastic solids are advanced functional materials that combine the elasticity of polymeric solids with the self-organisation of liquid crystalline structures. These materials were originally proposed by de Gennes (1975) \cite{deGennes:1975} and synthesized by Finkelmann et al. (1981) \cite{Finkelmann:1981:FKR}. As demonstrated by numerous experiments, their deformations can arise spontaneously and reversibly under certain external stimuli (heat, light, solvents), and are typically large and nonlinear \cite{Kundler:1995:KF,Kupfer:1991:KF,Kupfer:1994:KF,Zentel:1986}. This renders them useful as soft actuators \cite{White:2015:WB,deJeu:2012}. 

An ideal incompressible and microstructurally inextensible nematic solid is described by a strain-energy density function, depending only on the deformation gradient $\textbf{F}$  with respect to a reference configuration and the local nematic director $\textbf{n}$, such that $\det\textbf{F}=1$ (incompressibility) and $|\textbf{n}|=1$ (normality). For \textit{monodomain nematic solids},  where the director field does not vary from point to point, a well-known constitutive model is provided by the phenomenological \textit{neoclassical model}  developed in \cite{Bladon:1994:BTW,Warner:1988:WGV,Warner:1991:WW}. This model has been applied to predict large deformations in various applications \cite{Modes:2010:MBW,Modes:2011:MBW,Modes:2011:MW,Mostajeran:2016:MWWW,Warner:2018:WM}. In general, the neoclassical theory extends the molecular network theory of rubber elasticity \cite{Treloar:2005} to liquid crystal elastomers (LCE), whereby the constitutive parameters are directly measurable experimentally, or derived from macroscopic shape changes \cite{Warner:1996:WT,Warner:2007:WT}. 

However, similarly to the case of rubber elasticity, where ``no single value of modulus is applicable in the region of large strains'' \cite[p.~172]{Treloar:2005} (see \cite{Mihai:2017:MG} for a review), the neoclassical theory requires modifications in order to accommodate material behaviours observed in certain experiments \cite{Mao:1998:MWTB}. Therefore, various nonlinear expressions for the elastic contribution to the strain-energy function have been proposed, extending the neoclassical form \cite{Agostiniani:2012:ADS,Fried:2004:FS,Fried:2006:FS}. Of particular interest here are the governing equations developed in \cite{Fried:2004:FS,Fried:2006:FS}, which depend on two nonlinear  deformation tensors: the Cauchy-Green tensor that describes the overall macroscopic deformation, and a relative tensor that represents the microstructural deviation from the macroscopic deformation. As a special case, the model may contain, for example, the neo-Hookean and the neoclassical strain-energy densities. 

In addition, nematic solids may also exhibit variations in their macroscopic responses under mechanical tests, for instance  when different samples of a material are tested. For example, to account for ``compositional fluctuations'' causing variations in the local nematic responses (but not in the director orientation), in \cite{Verwey:1995:VW,Verwey:1996:VWT}, a small term was added to the one-term neoclassical form to obtain a mean value expression for the strain-energy function. 

In view of the aforementioned considerations, the aim of this study is twofold: first, to construct a class of generalised nematic constitutive models combining  elastic and neoclassical-type strain-energy functions, and second, to extend these elastic-nematic models to phenomenological stochastic continuum forms where the given parameters are random variables characterised by probability density functions. To attain this, we start with a  generalised nematic strain-energy function similar to that proposed by \cite{Fried:2004:FS,Fried:2006:FS}, where we take as reference configuration the stress-free state of the isotropic phase at high temperature \cite{Cirak:2014:CLBW,Conti:2002:CdSD,DeSimone:2000:dSD,DeSimone:2002:dSD,DeSimone:2009:dST}, instead of the nematic phase in which the cross-linking was produced \cite{Anderson:1999:ACF,Bladon:1993:BTW,Bladon:1994:BTW,Fried:2004:FS,Fried:2006:FS,Verwey:1996:VWT,Zhang:2019:etal}. Recognising the fact that some uncertainties may arise in the mechanical responses of liquid crystal elastomers, and inspired by recent developments in stochastic finite elasticity \cite{Fitt:2019:FWWM,Mihai:2019a:MDWG,Mihai:2019b:MDWG,Mihai:2019c:MDWG,Mihai:2018:MWG,Mihai:2019a:MWG,Mihai:2019b:MWG,Mihai:2020:MWG,Staber:2015:SG,Staber:2016:SG,Staber:2017:SG,Staber:2018:SG,Staber:2019:SGSMI}, we then design stochastic-elastic-nematic strain-energy functions to capture the variability in physical responses of nematic solid materials at a macroscopic level. 

To investigate the fundamental behaviour of the proposed elastic-nematic models and illustrate how the probabilistic parameters can be integrated into the nonlinear field theory, we apply these ideas to the classical problem of shear striping in a stretched nematic elastomer \cite{Conti:2002:CdSD,DeSimone:2000:dSD,DeSimone:2009:dST,Diaz:2012:DCR,Fried:2006:FS,Silhavy:2007,Verwey:1996:VWT,Zhang:2019:etal}, for which the solution can be derived explicitly. Early experimental studies of inhomogeneous shear stripes formation in stretched nematic elastomers were reported in \cite{Finkelmann:1997:FKTW,Kundler:1995:KF,Talroze:1999:etal,Zubarev:1999:etal}.

 Our analysis shows that, unlike in the purely neoclassical case, where the inhomogeneous deformation occurs within a constant interval that is independent of the model coefficients, for the elastic-nematic models, the critical interval depends on the ratio between the coefficients of the elastic and the neoclassical part. Moreover, for the stochastic form, the bounds of this interval are probabilistic, and the homogeneous and inhomogeneous states compete in the sense that both have a quantifiable chance to be obtained with a given probability. We call the inhomogeneous pattern developing within this interval `likely striping'. In Section~\ref{LCE:sec:models} , the governing equations for the elastic-nematic constitutive models are described , and the stochastic framework is presented and explained. This is followed in Section~\ref{LCE:sec:striping} by the analysis of shear striping in stretched nematic elastomers characterised by the new constitutive models in both deterministic and stochastic forms. Section~\ref{LCE:sec:conclude} contains our concluding remarks.

\section{Deterministic and stochastic models}\label{LCE:sec:models}

\subsection{Deterministic models}
\begin{figure}[htbp]
	\begin{center}
		\subfigure[]{\includegraphics[width=0.25\textwidth]{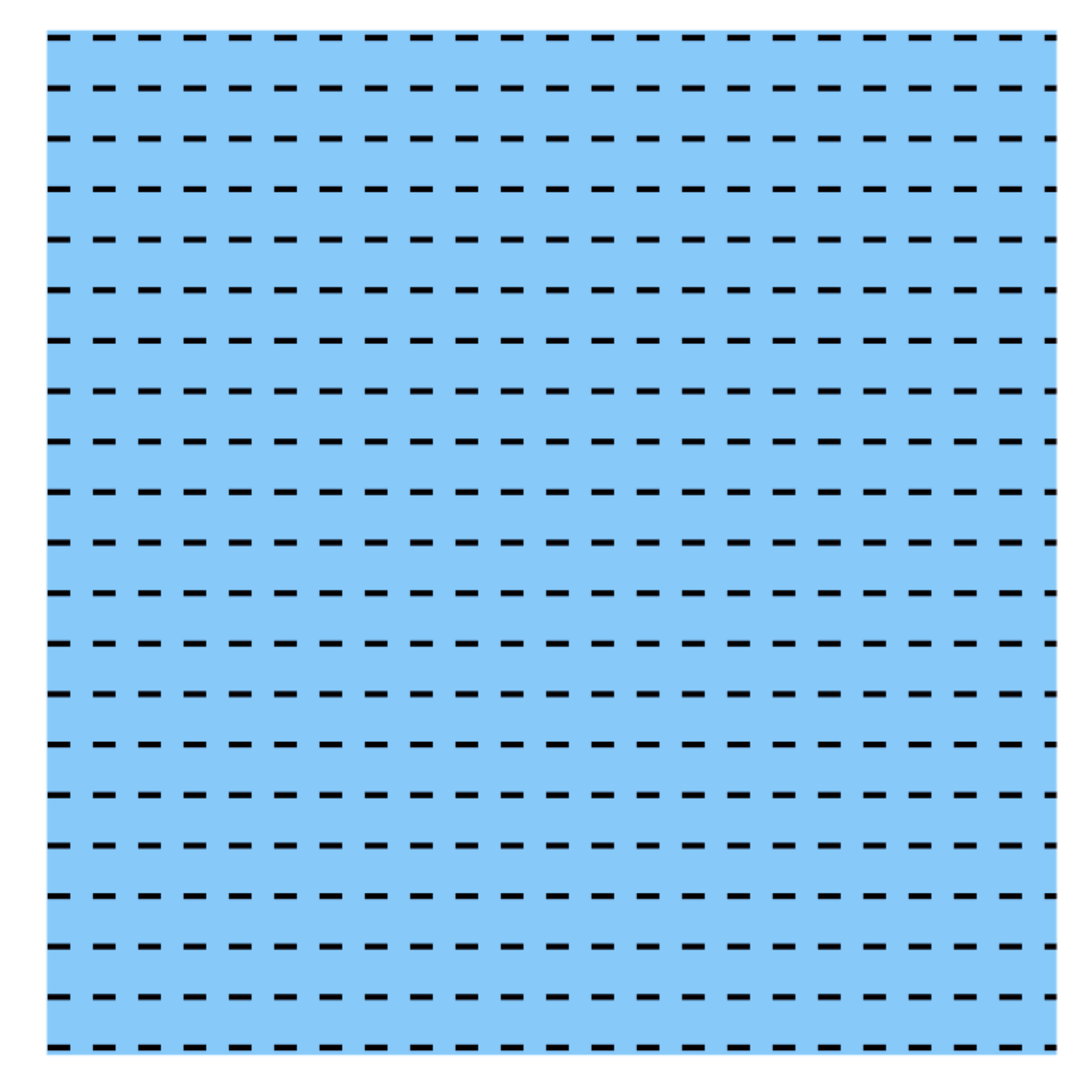}}\quad
		\subfigure[]{\includegraphics[width=0.7\textwidth]{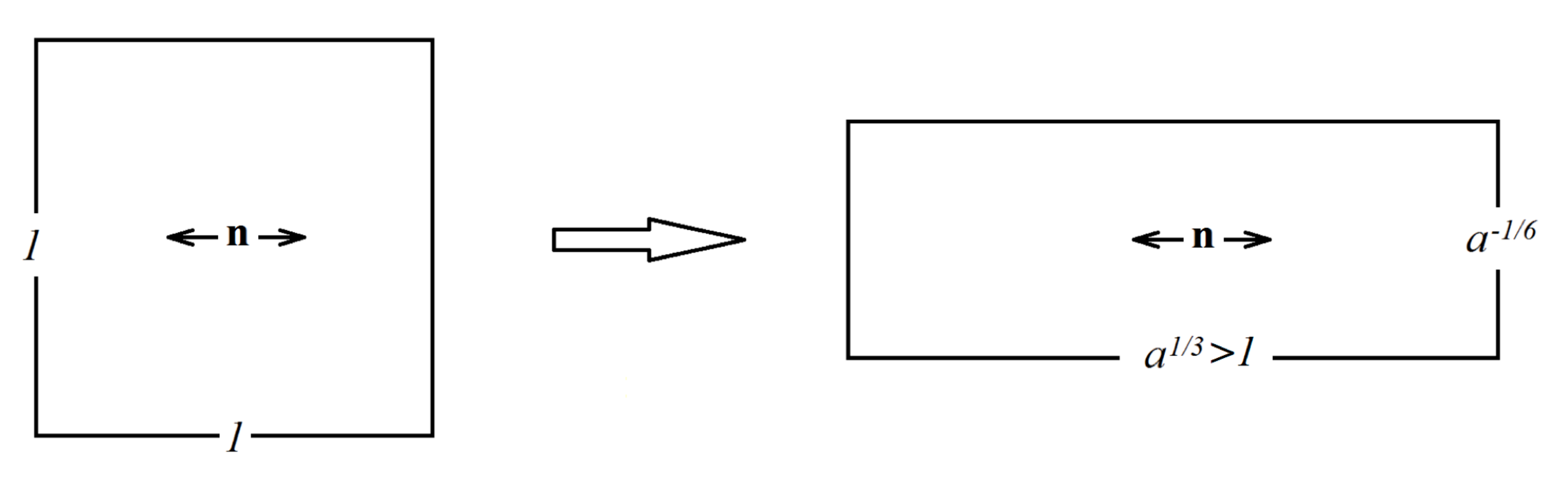}}
		\caption{Schematics of (a) nematic plate with uniform director field and (b) natural stretch when $a>1$.}\label{LCE:fig:nematic1}
	\end{center}
\end{figure}

The neoclassical strain-energy density function describing a nematic solid takes the general form
\begin{equation}\label{LCE:eq:Wnc:Fn}
W_{nc}(\textbf{F},\textbf{n})=W_{iso}(\textbf{F}^{e}),
\end{equation}
where $\textbf{F}$ is the deformation gradient from the isotropic state and $W_{iso}(\textbf{F}^{e})$ represents the strain-energy density of the isotropic polymer network depending only on the elastic deformation gradient $\textbf{F}^{e}$. The tensors $\textbf{F}$ and $\textbf{F}^{e}$ are related through
\begin{equation}\label{LCE:eq:F}
\textbf{F}=\textbf{G}\textbf{F}^{e},
\end{equation}
where
\begin{equation}\label{LCE:eq:G}
\textbf{G}=a^{1/3}\textbf{n}\otimes\textbf{n}+a^{-1/6}\left(\textbf{I}-\textbf{n}\otimes\textbf{n}\right)=a^{-1/6}\textbf{I}+\left(a^{1/3}-a^{-1/6}\right)\textbf{n}\otimes\textbf{n}
\end{equation} 
is the `spontaneous' gradient due to the liquid crystal director, with $a>0$ a temperature-dependent stretch parameter, $\otimes$ the tensor product of two vectors, and $\textbf{I}$ the identity tensor (see Figure~\ref{LCE:fig:nematic1}). In \eqref{LCE:eq:F}, $\textbf{G}$ is superimposed after the elastic deformation, defining a change of frame of reference from the isotropic phase to a nematic phase. Note that $\textbf{G}$ given by \eqref{LCE:eq:G} is symmetric, i.e., $\textbf{G}=\textbf{G}^{T}$ (with the superscript ``$T$'' denoting the transpose operation), whereas the deformation gradient $\textbf{F}^{e}$ may not be symmetric in general. In \eqref{LCE:eq:G}, $a^{1/3}$ and $a^{-1/6}$ are the stretch ratios parallel and perpendicular to the nematic director $\textbf{n}$, respectively. Their quotient, $r=a^{1/3}/a^{-1/6}=a^{1/2}$, represents the anisotropy parameter, which, in an ideal nematic solid, is the same in all directions. In the nematic phase, both the cases with $r>1$ (prolate shapes) and $r<1$ (oblate shapes) are possible. When $r=1$, the energy function reduces to that of an isotropic hyperelastic material. Typically, monodomains can be synthesised with the parameter $a$ taking values from $1.05$ (glasses) to $60$ (nematic rubber), which would correspond to changes in natural length between 7\% and 400\% (spontaneous extension ratio $a^{1/3}$ between $1.02$ and $4$) \cite{Clarke:2001:CHTT}. 

If $\textbf{n}_{0}$ is the reference orientation of the local director corresponding to the cross-linking state, then $\textbf{n}$ may differ from $\textbf{n}_{0}$ both by a rotation and a change in $r$. Here, we restrict our attention to imposed deformations without temperature changes. In this case the spontaneous gradient $\textbf{G}_{0}$ takes the form \eqref{LCE:eq:G} with $\textbf{n}_{0}$ instead of $\textbf{n}$.  

By \eqref{LCE:eq:G}, if $\textbf{R}$ is a rigid-body rotation, i.e., $\textbf{R}^{-1}=\textbf{R}^{T}$ and $\det\textbf{R}=1$, then the following identity holds
\begin{equation}\label{LCE:eq:G:Rn}
\textbf{R}^{T}\textbf{G}\textbf{R}=a^{-1/6}\textbf{I}+\left(a^{1/3}-a^{-1/6}\right)\left(\textbf{R}^{T}\textbf{n}\right)\otimes\left(\textbf{R}^{T}\textbf{n}\right).
\end{equation}
Since the nematic director $\textbf{n}$ is an observable (spatial) quantity, for any arbitrary reference orientation of the local director, $\textbf{n}_{0}$, there exists a rigid-body rotation $\textbf{Q}$, such that $\textbf{n}=\textbf{Q}\textbf{n}_{0}$. Hence,
\begin{equation}\label{LCE:eq:nQ}
\textbf{n}\otimes\textbf{n}=\textbf{Q}\left(\textbf{n}_{0}\otimes\textbf{n}_{0}\right)\textbf{Q}^{T}
\end{equation} 
and
\begin{equation}\label{LCE:eq:GG0}
\textbf{G}=\textbf{Q}\textbf{G}_{0}\textbf{Q}^{T}.
\end{equation} 

The neoclassical strain-energy function $W_{nc}$ given by \eqref{LCE:eq:Wnc:Fn} is minimised by any deformation satisfying $\textbf{F}\textbf{F}^{T}=\textbf{G}^2$, therefore, every pair $(\textbf{G}\textbf{R},\textbf{n})$, where $\textbf{R}$ is an arbitrary rigid-body rotation, is a natural (i.e., stress-free) state for this material model. 

A more general constitutive model that combines classical (standard) isotropic and neoclassical strain-energy densities can be defined as follows,
\begin{equation}\label{LCE:eq:W:Fn}
W(\textbf{F},\textbf{n})=W_{iso}\left(\textbf{F}\textbf{G}_{0}^{-1}\right)+W_{nc}(\textbf{F},\textbf{n}),
\end{equation}
where the first term can be viewed as the energy of the `parent' isotropic phase \cite{Conti:2002:CdSD,Fried:2004:FS}, and the second term is described by \eqref{LCE:eq:Wnc:Fn}. The function \eqref{LCE:eq:W:Fn} is minimised by every pair $(\textbf{G}_{r},\textbf{n})$ with $\textbf{G}_{r}=\textbf{G}\textbf{R}$, where $\textbf{G}$ is given by \eqref{LCE:eq:G} and $\textbf{R}$ is an arbitrary rigid-body rotation.

In general, a strain energy of the form \eqref{LCE:eq:Wnc:Fn} or \eqref{LCE:eq:W:Fn} relies on the following physically realistic assumptions. Note that these are also requirements for the elastic strain-energy density $W_{iso}$, and hence are inherited from isotropic finite elasticity \cite{goriely17,Mihai:2017:MG,Ogden:1997,TruesdellNoll:2004}:
\begin{itemize}
	\item[(A1)] Material objectivity, stating that constitutive equations must be invariant under changes of frame of reference. This requires that the scalar strain-energy function $W(\textbf{F},\textbf{n})$ is unaffected by a superimposed rigid-body transformation (which involves a change of position) after deformation, i.e., $W(\textbf{R}^{T}\textbf{F},\textbf{R}^{T}\textbf{n})=W(\textbf{F},\textbf{n})$, where $\textbf{R}\in SO(3)$ is a proper orthogonal tensor (rotation). Note that, as $\textbf{n}$ is defined with respect to the deformed configuration, it transforms when this configuration is rotated, whereas $\textbf{n}_{0}$ does not \cite{Fried:2004:FS}. Material objectivity is guaranteed by defining strain-energy functions in terms of the scalar invariants.
	
	Indeed, by the material frame indifference of $W_{iso}$, 
	\begin{equation}\label{LCE:eq:Wiso:objectivity}
	W_{iso}(\textbf{R}^{T}\textbf{F}^{e})=W_{iso}(\textbf{F}^{e}),
	\end{equation}
	and by \eqref{LCE:eq:F},
	\begin{equation}\label{LCE:eq:QF}
	\textbf{R}^{T}\textbf{F}=\left(\textbf{R}^{T}\textbf{G}\textbf{R}\right)\left(\textbf{R}^{T}\textbf{F}^{e}\right).
	\end{equation}
	Then, \eqref{LCE:eq:Wnc:Fn}, \eqref{LCE:eq:G:Rn}, \eqref{LCE:eq:Wiso:objectivity} and \eqref{LCE:eq:QF} imply
	\begin{equation}\label{LCE:eq:Wnc:objectivity}
	W_{nc}(\textbf{R}^{T}\textbf{F},\textbf{R}^{T}\textbf{n})=W_{iso}(\textbf{R}^{T}\textbf{F}^{e})=W_{iso}(\textbf{F}^{e})=W_{nc}(\textbf{F},\textbf{n}).
	\end{equation}
	Also, by the material frame indifference of $W_{iso}$, 
	\begin{equation}\label{LCE:eq:W:objectivity}
	W_{iso}\left(\textbf{R}^{T}\textbf{F}\textbf{G}_{0}^{-1}\right)=W_{iso}\left(\textbf{F}\textbf{G}_{0}^{-1}\right).
	\end{equation}
	
	\item[(A2)] Material isotropy, requiring that the strain-energy density is unaffected by a superimposed rigid-body transformation prior to deformation, i.e., $W(\textbf{F}\textbf{Q},\textbf{n})=W(\textbf{F},\textbf{n})$, where $\textbf{Q}\in SO(3)$. Note that, as $\textbf{n}$ is defined with respect to the deformed configuration, it does not change when the reference configuration is rotated, whereas $\textbf{n}_{0}$ does \cite{Fried:2004:FS}. For isotropic materials, the  strain-energy  function is a symmetric function of the principal stretch ratios $\{\lambda_{i}\}_{i=1,2,3}$ of $\textbf{F}$, i.e., $W(\textbf{F},\textbf{n})= W(\lambda_{1},\lambda_{2},\lambda_{3},\textbf{n})$. The squares of the principal stretches, $\{\lambda^2_{i}\}_{i=1,2,3}$, are the eigenvalues of the Cauchy-Green tensors, $\textbf{B}=\textbf{F}\textbf{F}^{T}$ and $\textbf{C}=\textbf{F}^{T}\textbf{F}$.
	
    This is because, as $W_{iso}$ is isotropic, i.e.,
	\begin{equation}\label{LCE:eq:Wiso:isotropy}
	W_{iso}(\textbf{F}^{e})=W_{iso}(\textbf{F}^{e}\textbf{Q}),
	\end{equation}
	and \eqref{LCE:eq:F} holds, it follows that
	\begin{equation}\label{LCE:eq:FQ}
	\textbf{F}\textbf{Q}=\textbf{G}\left(\textbf{F}^{e}\textbf{Q}\right).
	\end{equation}
	Hence, by \eqref{LCE:eq:Wnc:Fn}, \eqref{LCE:eq:Wiso:isotropy} and \eqref{LCE:eq:FQ},
	\begin{equation}\label{LCE:eq:Wnc:isotropy}
	W_{nc}(\textbf{F}\textbf{Q},\textbf{n})=W_{iso}(\textbf{F}^{e}\textbf{Q})=W_{iso}(\textbf{F}^{e})=W_{nc}(\textbf{F},\textbf{n}).
	\end{equation}
	Note that
	\begin{equation}\label{LCE:eq:W:isotropy}
	W_{iso}\left(\textbf{F}\textbf{Q}\left(\textbf{G}_{0}\textbf{Q}\right)^{-1}\right)=W_{iso}\left(\textbf{F}\textbf{G}_{0}^{-1}\right).
	\end{equation}
	
	\item[(A3)] Baker-Ericksen (BE) inequalities, stating that \emph{the greater principal (Cauchy) stress occurs in the direction of the greater principal elastic stretch} \cite{BakerEricksen:1954,Silhavy:2007}. These inequalities take the form
	\begin{equation}\label{LCE:eq:Wnc:BE}
	\left(\lambda_{i}\frac{\partial W_{nc}}{\partial\lambda_{i}}-\lambda_{j}\frac{\partial W_{nc}}{\partial\lambda_{j}}\right)\left(\lambda_{i}-\lambda_{j}\right)>0\quad \mbox{if}\quad \lambda_{i}\neq\lambda_{j},\quad i,j=1,2,3.
	\end{equation}
	When any two principal stretches are equal, the strict inequality ``$>$'' in \eqref{LCE:eq:Wnc:BE} is replaced by ``$\geq$''.
\end{itemize}

In view of assumption (A1), the neoclassical model defined by \eqref{LCE:eq:Wnc:Fn} can be expressed equivalently in the form
\begin{equation}\label{LCE:eq:Wnc:I12345}
 W_{nc}(I_{1},I_{2},I_{3},I_{4},I_{5})= W_{iso}(I_{1}^{e},I_{2}^{e},I_{3}^{e}),
\end{equation}
where 
\begin{eqnarray}
&&I_{1}=\text{tr}\left(\textbf{F}\textbf{F}^{T}\right)=\text{tr}\left(\textbf{F}^{T}\textbf{F}\right),\label{LCE:eq:I1}\\	&&I_{2}=\text{tr}\left[\text{Cof}\left(\textbf{F}\textbf{F}^{T}\right)\right]=\text{tr}\left[\text{Cof}\left(\textbf{F}^{T}\textbf{F}\right)\right],\label{LCE:eq:I2}\\
&&I_{3}=\det\left(\textbf{F}\textbf{F}^{T}\right)=\det\left(\textbf{F}^{T}\textbf{F}\right),\label{LCE:eq:I3}\\
&&I_{4}=\textbf{n}\cdot\textbf{F}\textbf{F}^{T}\textbf{n},\label{LCE:eq:I4}\\
&&I_{5}=\textbf{n}\cdot\left(\textbf{F}\textbf{F}^{T}\right)^2\textbf{n}\label{LCE:eq:I5}
\end{eqnarray}
are the scalar invariants, with $\text{tr}(\textbf{A})$ representing the trace and $\text{Cof}\left(\textbf{A}\right)=\det(\textbf{A})\textbf{A}^{-T}$ denoting the cofactor of a tensor $\textbf{A}$, while $ W_{iso}$ depends only on the principal invariants $I_{1}^{e},I_{2}^{e},I_{3}^{e}$, which take the forms \eqref{LCE:eq:I1}-\eqref{LCE:eq:I3} with $\textbf{F}^{e}$ instead of $\textbf{F}$. An invariant formulation of strain-energy density functions is a powerful tool of solid mechanics \cite{goriely17,Mihai:2017:MG,Ogden:1997}. Yet, to the best of our knowledge, it does not appear to have been used systematically in the theory of nematic elastomers, but only briefly mentioned in \cite{Fried:2004:FS}.

By \eqref{LCE:eq:F} and \eqref{LCE:eq:G}, the following relations hold between the invariants $\{I_{1},I_{2},I_{3},I_{4},I_{5}\}$  and $\{I_{1}^{e},I_{2}^{e},I_{3}^{e}\}$:
\begin{eqnarray}
&& I_{1}^{e}=a^{1/3}\left[I_{1}-\left(1- a^{-1}\right)I_{4}\right],\label{LCE:eq:I1e}\\
&& I_{2}^{e}=a^{-1/3}\left[I_{2}-\left(1-a\right)I_{3}^{-1}\left(I_{5}-I_{1}I_{4}+I_{2}\right)\right],\label{LCE:eq:I2e}\\
&& I_{3}^{e}=I_{3}.\label{LCE:eq:I3e}
\end{eqnarray}
Clearly, identity \eqref{LCE:eq:I3e} is valid since $\det \textbf{G}=1$, and therefore
\begin{equation}\label{LCE:eq:I3e:proof}
I_{3}^{e}=\left(\det \textbf{F}^{e}\right)^2=\left(\det \textbf{G}^{-1}\det \textbf{F}\right)^2=\left(\det \textbf{F}\right)^2=I_{3}.
\end{equation}
Identity \eqref{LCE:eq:I1e} holds since
\begin{equation}\label{LCE:eq:I1e:proof}
\begin{split}
I_{1}^{e}&=\text{tr}\left(\textbf{F}^{eT}\textbf{F}^{e}\right)\\
&=\text{tr}\left[\textbf{F}^{T}\textbf{G}^{-2}\textbf{F}\right]\\
&=a^{1/3}\left[\text{tr}\left(\textbf{F}^{T}\textbf{F}\right)-\left(1- a^{-1}\right)\textbf{n}\cdot\textbf{F}\textbf{F}^{T}\textbf{n}\right]\\
&=a^{1/3}\left[I_{1}-\left(1- a^{-1}\right)I_{4}\right].
\end{split}
\end{equation}
To show identity \eqref{LCE:eq:I2e}, we recall the Cayley-Hamilton theorem that
\begin{equation}\label{LCE:eq:CH}
\left(\textbf{F}\textbf{F}^{T}\right)^3-I_{1}\left(\textbf{F}\textbf{F}^{T}\right)^2+I_{2}\left(\textbf{F}\textbf{F}^{T}\right)-I_{3}\textbf{I}=0.
\end{equation}
Multiplying the above equation by $\left(\textbf{F}\textbf{F}^{T}\right)^{-1}$ implies
\begin{equation}\label{LCE:eq:CH:inverse}
I_{3}\left(\textbf{F}\textbf{F}^{T}\right)^{-1}=\left(\textbf{F}\textbf{F}^{T}\right)^2-I_{1}\left(\textbf{F}\textbf{F}^{T}\right)+I_{2}\textbf{I}.
\end{equation}
Hence,
\begin{equation}\label{LCE:eq:I2e:CH:n}
\textbf{n}\cdot\textbf{F}^{-T}\textbf{F}^{-1}\textbf{n}
=\textbf{n}\cdot I_{3}^{-1}\left[\left(\textbf{F}\textbf{F}^{T}\right)^2-I_{1}\left(\textbf{F}\textbf{F}^{T}\right)+I_{2}\textbf{I}\right]\textbf{n}\\
=I_{3}^{-1}\left(I_{5}-I_{1}I_{4}+I_{2}\right).
\end{equation}
Finally, by calculations analogous to those in \eqref{LCE:eq:I1e:proof}, together with \eqref{LCE:eq:I3e} and \eqref{LCE:eq:I2e:CH:n}, it follows that 
\begin{equation}\label{LCE:eq:I2e:proof}
\begin{split}
I_{2}^{e}&=I_{3}^{e}\text{tr}\left(\textbf{F}^{eT}\textbf{F}^{e}\right)^{-T}\\
&=I_{3}\text{tr}\left[\textbf{F}^{-1}\textbf{G}^{2}\textbf{F}^{-T}\right]\\
&=I_{3}a^{-1/3}\left[\text{tr}\left(\textbf{F}^{-1}\textbf{F}^{-T}\right)-\left(1- a\right)\textbf{n}\cdot\textbf{F}^{-T}\textbf{F}^{-1}\textbf{n}\right]\\
&=a^{-1/3}\left[I_{2}-\left(1-a\right)\left(I_{5}-I_{1}I_{4}+I_{2}\right)\right].
\end{split}
\end{equation}

Assumption (A2) implies that the nematic model given by \eqref{LCE:eq:Wnc:Fn} can the written equivalently in the form
\begin{equation}\label{LCE:eq:Wnc:lambda123}
 W_{nc}(\lambda_{1},\lambda_{2},\lambda_{3},\textbf{n})= W_{iso}(\lambda_{1}^{e},\lambda_{2}^{e},\lambda_{3}^{e}),
\end{equation}
where $\lambda_{1}, \lambda_{2}, \lambda_{3}$ are the principal stretches of the LCE, and $\lambda_{1}^{e}, \lambda_{2}^{e}, \lambda_{3}^{e}$ are the principal stretches for the corresponding polymeric network. Then, $\{\lambda_{1},\lambda_{2},\lambda_{3},\textbf{n}\}$  and $\{\lambda_{1}^{e},\lambda_{2}^{e},\lambda_{3}^{e}\}$ are related though the identities \eqref{LCE:eq:I1e}-\eqref{LCE:eq:I3e}, the spectral decomposition of $\textbf{F}\textbf{F}^{T}$, and the usual formulae for the principal invariants in terms of the principal stretches, i.e.,
\begin{equation}\label{LCE:eq:lambdaI123}
\begin{split}
I_{1}&=\lambda_{1}^2+\lambda_{2}^2+\lambda_{3}^2,\\
I_{2}&=\lambda_{1}^2\lambda_{2}^2+\lambda_{2}^2\lambda_{3}^2+\lambda_{3}^2\lambda_{1}^2,\\
I_{3}&=\lambda_{1}^2\lambda_{2}^2\lambda_{3}^2,
\end{split}
\end{equation}
and
\begin{equation}\label{LCE:eq:lambdaI123e}
\begin{split}
I_{1}^{e}&=\left(\lambda_{1}^{e}\right)^2+\left(\lambda_{2}^{e}\right)^2+\left(\lambda_{3}^{e}\right)^2,\\
I_{2}^{e}&=\left(\lambda_{1}^{e}\right)^2\left(\lambda_{2}^{e}\right)^2+\left(\lambda_{2}^{e}\right)^2\left(\lambda_{3}^{e}\right)^2+\left(\lambda_{3}^{e}\right)^2\left(\lambda_{3}^{e}\right)^2\\
I_{3}^{e}&=\left(\lambda_{1}^{e}\right)^2\left(\lambda_{2}^{e}\right)^2\left(\lambda_{3}^{e}\right)^2.
\end{split}
\end{equation}

Assumption (A3) ensures that the shear modulus is always positive \cite{Mihai:2017:MG}.

\subsection{Stochastic models}

In addition to the model assumptions (A1)-(A3) presented in Section~\ref{LCE:sec:models}, we adopt the following hypothesis which is similar to that required by recent stochastic hyperelastic models analysed in \cite{Fitt:2019:FWWM,Mihai:2019a:MDWG,Mihai:2019b:MDWG,Mihai:2019c:MDWG,Mihai:2018:MWG, Mihai:2019a:MWG,Mihai:2020:MWG}:
\begin{itemize}
\item[(A4)] For any given finite deformation, at any point in the material, the shear modulus, $\mu$, and its inverse, $1/\mu$, are second-order random variables, i.e., they have finite mean value and finite variance \cite{Staber:2015:SG,Staber:2016:SG,Staber:2017:SG,Staber:2018:SG,Staber:2019:SGSMI}.
\end{itemize}
This assumption is guaranteed by setting the following mathematical expectations:
\begin{eqnarray}\label{eq:Emu1}\begin{cases}
E\left[\mu\right]=\underline{\mu}>0,&\\
E\left[\log\ \mu\right]=\nu,& \mbox{such that $|\nu|<+\infty$}.\label{eq:Emu2}\end{cases}
\end{eqnarray}
The first constraint in \eqref{eq:Emu1} specifies the mean value for the random shear modulus $\mu$, while the second constraint provides a condition from which it follows that $1/\mu$ is a second order random variable \cite{Soize:2001}, \cite[p.~270]{Soize:2017}. Then, by the maximum entropy principle \cite{Jaynes:2003}, the shear modulus $\mu$ with mean value $\underline{\mu}$ and standard deviation $\|\mu\|=\sqrt{\text{Var}[\mu]}$ (defined as the square root of the variance, $\text{Var}[\mu]$) follows a Gamma probability distribution \cite{Soize:2000,Soize:2001,Soize:2006,Soize:2017} with shape and scale parameters $\rho_{1}>0$ and $\rho_{2}>0$ respectively, such that
\begin{equation}\label{eq:rho12}
\underline{\mu}=\rho_{1}\rho_{2},\qquad
\|\mu\|=\sqrt{\rho_{1}}\rho_{2}.
\end{equation}
The corresponding probability density function takes the form \cite{Abramowitz:1964,Grimmett:2001:GS,Johnson:1994:JKB}
\begin{equation}\label{eq:mu:gamma}
g(\mu;\rho_{1},\rho_{2})=\frac{\mu^{\rho_{1}-1}e^{-\mu/\rho_{2}}}{\rho_{2}^{\rho_{1}}\Gamma(\rho_{1})},\qquad\mbox{for}\ \mu>0\ \mbox{and}\ \rho_{1}, \rho_{2}>0,
\end{equation}
where $\Gamma:\mathbb{R}^{*}_{+}\to\mathbb{R}$ is the complete Gamma function
\begin{equation}\label{eq:gamma}
\Gamma(z)=\int_{0}^{+\infty}t^{z-1}e^{-t}\text dt.
\end{equation}
The word `hyperparameters' is sometimes used for $\rho_{1}$ and $\rho_{2}$ to distinguish them from $\mu$ and other material constants \cite[p.~8]{Soize:2017}.

\begin{figure}[htbp]
	\begin{center}
		\includegraphics[width=0.6\textwidth]{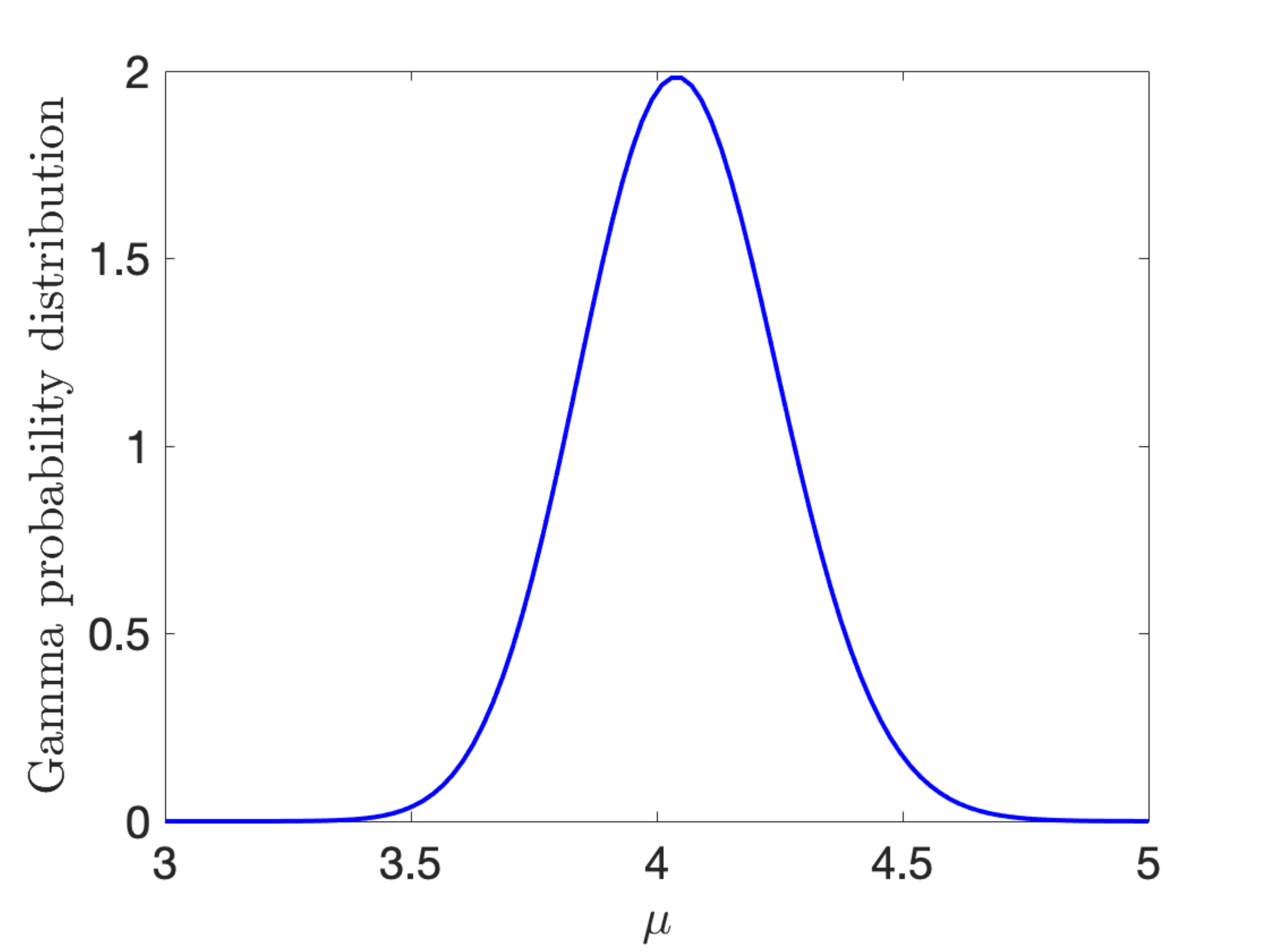}
		\caption{Example of Gamma distribution, defined by \eqref{eq:mu:gamma}, with shape and scale parameters $\rho_{1}=405$, $\rho_{2}=0.01$, respectively.}\label{fig:mu-gpdf}
	\end{center}
\end{figure}

For illustration, a shear modulus $\mu$ satisfying the Gamma distribution with hyperparameters $\rho_{1}=405$ and $\rho_{2}=0.01$ is shown in Figure~\ref{fig:mu-gpdf}. In this case, there is a strong similarity between the Gamma and a normal distributions, due to the fact that $\rho_{1}$ is much larger than $\rho_{2}$ (see \cite{Fitt:2019:FWWM,Mihai:2019a:MDWG}). However, elastic moduli cannot be described by the normal distribution, since this distribution is defined on the entire real line whereas elastic moduli are generally positive. In practice, these moduli can meaningfully take on different values corresponding to possible outcomes of experimental tests. The principle of maximum entropy then enables the explicit construction of their probability distributions based on the available information. If the data cannot be smaller than a certain bound different from zero, then a shift parameter can be incorporated to define a three-parameter Gamma distribution. Approaches for the explicit derivation of probability distributions for the elastic parameters of stochastic homogeneous isotropic hyperelastic models calibrated to experimental data for rubber-like material and soft tissues are presented in \cite{Mihai:2018:MWG,Staber:2017:SG}. 

\subsection{Mooney-Rivlin-type stochastic nematic models} 

The classical Mooney-Rivlin (MR) model \cite{Mooney:1940,Rivlin:1948} captures well the elastic behaviour of polymeric networks at small and medium strains \cite{Mihai:2017:MG,Ogden:2004:OSS,Pucci:2002:PS}. Here, we consider the stochastic version \cite{Mihai:2019a:MWG}
\begin{equation}\label{LCE:eq:Wiso:MR}
W_{iso}(\textbf{F}^{e})=\frac{\mu_{1}}{2}\left[\text{tr}\left(\textbf{F}^{e}\textbf{F}^{eT}\right)-3\right]+\frac{\mu_{2}}{2}\left\{\text{tr}\left[\text{Cof}\left(\textbf{F}^{e}\textbf{F}^{eT}\right)\right]-3\right\},
\end{equation}
where $\mu_{1}+\mu_{2}>0$ is the shear modulus for infinitesimal deformations, with $\mu_{1}$ and $\mu_{2}$ random variables. When $\mu_{2}=0$, the MR model reduces to the neo-Hookean (NH) form \cite{Treloar:1944}.

The neoclassical MR-type strain-energy function, given by \eqref{LCE:eq:Wnc:Fn}, takes the form \cite{Fried:2006:FS,Silhavy:2007},
\begin{equation}\label{LCE:eq:Wnc:Fn:MR}
\begin{split}
W_{nc}(\textbf{F},\textbf{n})&=\frac{\mu_{1}}{2}\left\{a^{1/3}\left[\text{tr}\left(\textbf{F}\textbf{F}^{T}\right)-\left(1- a^{-1}\right)\textbf{n}\cdot\textbf{F}\textbf{F}^{T}\textbf{n}\right]\right\}\\
&+\frac{\mu_{2}}{2}\left\{a^{-1/3}\left[\text{tr}\left(\textbf{F}^{-T}\textbf{F}^{-1}\right)-\left(1-a\right)\textbf{n}\cdot\textbf{F}^{-T}\textbf{F}^{-1}\textbf{n}\right]\right\}.
\end{split}
\end{equation}
This function is conveniently expressed  in terms of the invariants defined by  \eqref{LCE:eq:I1}-\eqref{LCE:eq:I5}, with $I_{3}=1$, as follows,
\begin{equation}\label{LCE:eq:Wnc:Is:MR}
\begin{split}
 W_{nc}(I_{1},I_{2},I_{4},I_{5})&=\frac{\mu_{1}}{2}a^{1/3}\left[I_{1}-\left(1- a^{-1}\right)I_{4}\right]\\
&+\frac{\mu_{2}}{2}a^{-1/3}\left[I_{2}-\left(1-a\right)\left(I_{5}-I_{1}I_{4}+I_{2}\right)\right].
\end{split}
\end{equation}
Equivalently, in terms of the principal stretches, this strain-energy function is given by
\begin{equation}\label{LCE:eq:Wnc:MR}
\begin{split}
 W_{nc}(\lambda_{1},\lambda_{2},\lambda_{3},\textbf{n})
&=\frac{\mu_{1}}{2}a^{1/3}\left[\sum_{i=1}^3\lambda_{i}^2-\left(1- a^{-1}\right)\sum_{i=1}^{3}\lambda_{i}^2\left(\textbf{e}_{i}\cdot\textbf{n}\right)^2\right]\\
&+\frac{\mu_{2}}{2}a^{-1/3}\left[\sum_{i=1}^3\lambda_{i}^{-2}-\left(1- a\right)\sum_{i=1}^{3}\lambda_{i}^{-2}\left(\textbf{e}_{i}\cdot\textbf{n}\right)^2\right].
\end{split}
\end{equation}

More generally, a nematic constitutive model combining classical and neoclassical MR-type strain-energy densities takes the form
\begin{equation}\label{LCE:eq:W:Fn:MR}
\begin{split}
W(\textbf{F},\textbf{n})
&=\frac{\mu^{(1)}_{1}}{2}\text{tr}\left(\textbf{F}\textbf{G}_{0}^{-2}\textbf{F}^{T}\right)\\
&+\frac{\mu^{(1)}_{2}}{2}\text{tr}\left(\textbf{F}^{-T}\textbf{G}_{0}^{2}\textbf{F}^{-1}\right)\\
&+\frac{\mu^{(2)}_{1}}{2}a^{1/3}\left[\text{tr}\left(\textbf{F}\textbf{F}^{T}\right)-\left(1- a^{-1}\right)\textbf{n}\cdot\textbf{F}\textbf{F}^{T}\textbf{n}\right]\\
&+\frac{\mu^{(2)}_{2}}{2}a^{-1/3}\left[\text{tr}\left(\textbf{F}^{-T}\textbf{F}^{-1}\right)-\left(1-a\right)\textbf{n}\cdot\textbf{F}^{-T}\textbf{F}^{-1}\textbf{n}\right],
\end{split}
\end{equation}
where the shear modulus under infinitesimal deformation is $\mu=\mu^{(1)}+\mu^{(2)}=\mu^{(1)}_{1}+\mu^{(1)}_{2}+\mu^{(2)}_{1}+\mu^{(2)}_{2}>0$, with $\mu^{(1)}=\mu^{(1)}_{1}+\mu^{(1)}_{2}>0$ and $\mu^{(2)}=\mu^{(2)}_{1}+\mu^{(2)}_{2}>0$.

\subsection{Gent-Gent-type stochastic nematic models} 

It is known that the Gent-Gent (GG) hyperelastic model reasonably captures  the nonlinear behaviour of polymeric networks at large strains \cite{Destrade:2017:DSS,Mihai:2017:MG,Ogden:2004:OSS,Pucci:2002:PS}. Its  stochastic version is given by
\begin{equation}\label{LCE:eq:Wiso:GG}
W_{iso}(\textbf{F}^{e})=-\frac{\mu_{1}}{2\beta}\ln\left\{1-\beta\left[\text{tr}\left(\textbf{F}^{e}\textbf{F}^{eT}\right)-3\right]\right\}+\frac{3\mu_{2}}{2}\ln\frac{\text{tr}\left[\text{Cof}\left(\textbf{F}^{e}\textbf{F}^{eT}\right)\right]}{3},
\end{equation}
where $\mu_{1}+\mu_{2}>0$ is the shear modulus in infinitesimal deformation, with $\mu_{1}$ and $\mu_{2}$ random variables, and $\beta$ is a deterministic constant independent of the deformation. If $\mu_{2}=0$, then the GG model reduces to the Gent form \cite{Gent:1996}.

The neoclassical GG strain-energy function, defined by \eqref{LCE:eq:Wnc:Fn}, takes the form
\begin{equation}\label{LCE:eq:Wnc:Fn:GG}
\begin{split}
W_{nc}(\textbf{F},\textbf{n})
&=-\frac{\mu_{1}}{2\beta}\ln\left\{1-\beta a^{1/3}\left[\text{tr}\left(\textbf{F}\textbf{F}^{T}\right)-\left(1- a^{-1}\right)\textbf{n}\cdot\textbf{F}\textbf{F}^{T}\textbf{n}\right]+3\beta\right\}\\
&+\frac{3\mu_{2}}{2}\ln\frac{a^{-1/3}\left[\text{tr}\left(\textbf{F}^{-T}\textbf{F}^{-1}\right)-\left(1-a\right)\textbf{n}\cdot\textbf{F}^{-T}\textbf{F}^{-1}\textbf{n}\right]}{3}.
\end{split}
\end{equation}
Equivalently, this function has the following expression depending on the invariants given by  \eqref{LCE:eq:I1}-\eqref{LCE:eq:I5}, with $I_{3}=1$, 
\begin{equation}\label{LCE:eq:Wnc:Is:GG}
\begin{split}
 W_{nc}(I_{1},I_{2},I_{4},I_{5})
&=-\frac{\mu_{1}}{2\beta}\ln\left\{1-\beta a^{1/3}\left[I_{1}-\left(1- a^{-1}\right)I_{4}\right]+3\beta\right\}\\
&+\frac{3\mu_{2}}{2}\ln\frac{a^{-1/3}\left[I_{2}-\left(1-a\right)\left(I_{5}-I_{1}I_{4}+I_{2}\right)\right]}{3}.
\end{split}
\end{equation}
In terms of the principal stretches, this strain-energy function is equivalent to
\begin{equation}\label{LCE:eq:Wnc:GG}
\begin{split}
 W_{nc}(\lambda_{1},\lambda_{2},\lambda_{3},\textbf{n})
&=-\frac{\mu_{1}}{2\beta}\ln\left\{1-\beta a^{1/3}\left[\sum_{i=1}^3\lambda_{i}^2-\left(1- a^{-1}\right)\sum_{i=1}^{3}\lambda_{i}^2\left(\textbf{e}_{i}\cdot\textbf{n}\right)^2\right]+3\beta\right\}\\
&+\frac{3\mu_{2}}{2}\ln\frac{a^{-1/3}\left[\sum_{i=1}^3\lambda_{i}^{-2}-\left(1- a\right)\sum_{i=1}^{3}\lambda_{i}^{-2}\left(\textbf{e}_{i}\cdot\textbf{n}\right)^2\right]}{3}.
\end{split}
\end{equation}

A nematic model that combines classical and neoclassical GG-type strain-energy densities is described by
\begin{equation}\label{LCE:eq:W:Fn:GG}
\begin{split}
W(\textbf{F},\textbf{n})
&=-\frac{\mu^{(1)}_{1}}{2\beta^{(1)}}\ln\left\{1-\beta^{(1)}\left[\text{tr}\left(\textbf{F}\textbf{G}_{0}^{-2}\textbf{F}^{T}\right)-3\right]\right\}+\frac{3\mu^{(1)}_{2}}{2}\ln\frac{\text{tr}\left(\textbf{F}^{-T}\textbf{G}_{0}^{2}\textbf{F}^{-1}\right)}{3}\\
&-\frac{\mu^{(2)}_{1}}{2\beta^{(2)}}\ln\left\{1-\beta^{(2)} a^{1/3}\left[\text{tr}\left(\textbf{F}\textbf{F}^{T}\right)-\left(1- a^{-1}\right)\textbf{n}\cdot\textbf{F}\textbf{F}^{T}\textbf{n}\right]+3\beta^{(2)}\right\}\\
&+\frac{3\mu^{(2)}_{2}}{2}\ln\frac{a^{-1/3}\left[\text{tr}\left(\textbf{F}^{-T}\textbf{F}^{-1}\right)-\left(1-a\right)\textbf{n}\cdot\textbf{F}^{-T}\textbf{F}^{-1}\textbf{n}\right]}{3},
\end{split}
\end{equation}
where $\mu=\mu^{(1)}+\mu^{(2)}=\mu^{(1)}_{1}+\mu^{(1)}_{2}+\mu^{(2)}_{1}+\mu^{(2)}_{2}>0$ is the shear modulus under small strains, such that $\mu^{(1)}=\mu^{(1)}_{1}+\mu^{(1)}_{2}>0$ and $\mu^{(2)}=\mu^{(2)}_{1}+\mu^{(2)}_{2}>0$, and $\beta^{(1)}$, $\beta^{(2)} $ are a given constants independent of the deformation.

\section{Shear striping in biaxial stretch}\label{LCE:sec:striping}

\begin{figure}[htbp]
	\begin{center}
	    \includegraphics[width=0.95\textwidth]{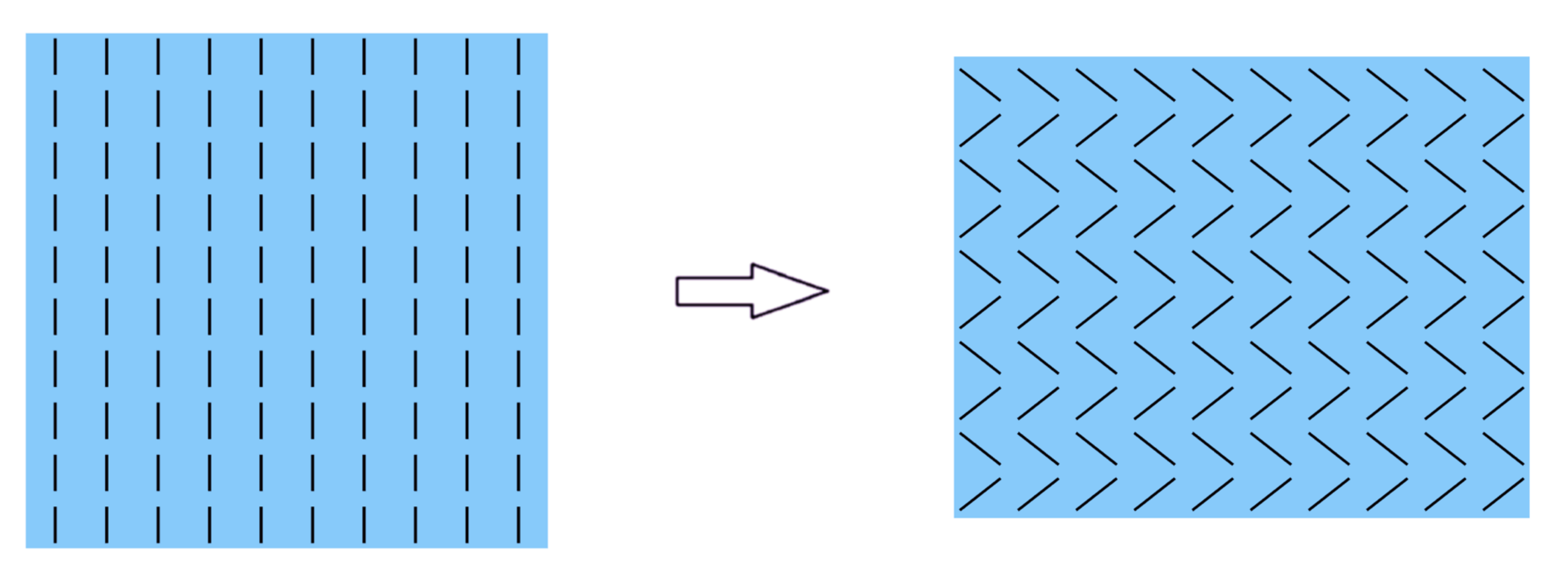}
		\caption{Schematics of nematic solid with uniform director field (left) presenting shear stripes during biaxial stretch (right).}\label{LCE:fig:nematic-stripes}
	\end{center}
\end{figure}

Many deformations of nematic polymers result in director rotations, and, in general, a uniform order, as depicted in Figure~\ref{LCE:fig:nematic-stripes}-left, will give rise to a uniform response. However, non-uniform behaviours are also possible. In particular, under biaxial (or ``pure shear'' \cite{Rivlin:1951:VII:RS}) stretch, an alternating simple shear striping develops during the gradual rotation of the nematic director from the initial perpendicular direction to the final parallel direction to the elongation \cite{Conti:2002:CdSD,DeSimone:2000:dSD,DeSimone:2009:dST,Fried:2006:FS}. This is illustrated in Figure~\ref{LCE:fig:nematic-stripes}-right. The relationship between biaxial stretch and simple shear deformations in large strain analysis is discussed in Appendix~\ref{LCE:sec:append}. The  explanation of this phenomenon is that, for these materials, the energy is minimised by passing through a state exhibiting a microstructure of many homogeneously deformed parts.  A characteristic aspect of these patterns is that they only exist in an interval of applied tension. A natural question is then: What is the influence of the constitutive model on the `window' of deformations within which striped domains are formed in a nematic elastomer under biaxial stretch? 

We set the Cartesian system of reference, such that \cite{DeSimone:2009:dST}
\begin{equation}\label{LCE:eq:n0ntheta}
\textbf{n}_{0}=
\left[
\begin{array}{c}
0\\
0\\
1
\end{array}
\right],\qquad
\textbf{n}=
\left[
\begin{array}{c}
0\\
\sin\theta\\
\cos\theta
\end{array}
\right],
\end{equation}
where $\theta$ is the angle between $\textbf{n}$ and $\textbf{n}_{0}$, and assume a deformation with gradient tensor \cite{DeSimone:2009:dST}
\begin{equation}\label{LCE:F:stretch}
\textbf{F}=
\left[
\begin{array}{ccc}
a^{-1/6} & 0 & 0\\
0 & \lambda & \varepsilon\\
0 & 0 & a^{1/6}\lambda^{-1}
\end{array}
\right],
\end{equation}
where $a>1$, $\lambda>0$, and $\varepsilon>0$. Note that, during this deformation, the material is stretched in the second direction, while in the first direction, it remains fixed. The corresponding left Cauchy-Green tensor and its cofactor are, respectively, 
\begin{eqnarray}
&&\textbf{B}=\textbf{F}\textbf{F}^{T}=
\left[
\begin{array}{ccc}
a^{-1/3} & 0 & 0\\
0 & \lambda^2+\varepsilon^2 & \varepsilon a^{1/6}\lambda^{-1}\\
0 & \varepsilon a^{1/6}\lambda^{-1} & a^{1/3}\lambda^{-2}
\end{array}
\right],\label{LCE:B:stretch}\\
&&\text{Cof}\ \textbf{B}=\textbf{F}^{-T}\textbf{F}^{-1}=
\left[
\begin{array}{ccc}
a^{1/3} & 0 & 0\\
0 & \lambda^{-2} & -\varepsilon a^{-1/6}\lambda^{-1}\\
0 & -\varepsilon a^{-1/6}\lambda^{-1} & a^{-1/3}\left(\lambda^2+\varepsilon^2\right)
\end{array}
\right].\label{LCE:invB:stretch}
\end{eqnarray}
It follows that
\begin{eqnarray}
&&\text{tr}\left(\textbf{F}\textbf{F}^{T}\right)=a^{-1/3}+\lambda^2+\varepsilon^2+a^{1/3}\lambda^{-2},\\
&&\textbf{n}\cdot\textbf{F}\textbf{F}^{T}\textbf{n}=\left(\lambda^2+\varepsilon^2\right)\sin^2\theta+\varepsilon a^{1/6}\lambda^{-1}\sin\left(2\theta\right)+a^{1/3}\lambda^{-2}\cos^2\theta,\\
&&\text{tr}\left(\textbf{F}^{-T}\textbf{F}^{-1}\right)=a^{1/3}+\lambda^{-2}+a^{-1/3}\left(\lambda^2+\varepsilon^2\right),\\
&&\textbf{n}\cdot\textbf{F}^{-T}\textbf{F}^{-1}\textbf{n}=\lambda^{-2}\sin^2\theta-\varepsilon a^{-1/6}\lambda^{-1}\sin\left(2\theta\right)+a^{-1/3}\left(\lambda^2+\varepsilon^2\right)\cos^2\theta.
\label{LCE:eq:traces:stretch}
\end{eqnarray}
Given the deformation gradient $\textbf{F}$ defined by \eqref{LCE:F:stretch} and the reference nematic director $\textbf{n}_{0}$ specified in \eqref{LCE:eq:n0ntheta}, we have
\begin{equation}\label{LCE:FG:stretch}
\textbf{F}\textbf{G}_{0}^{-1}=
\left[
\begin{array}{ccc}
1 & 0 & 0\\
0 & a^{1/6}\lambda &  a^{-1/3}\varepsilon\\
0 & 0 & a^{-1/6}\lambda^{-1}
\end{array}
\right].
\end{equation}

When a nematic model comprises both standard isotropic and neoclassical terms, the minimiser  depends on the material parameters  \cite{Fried:2004:FS,Fried:2006:FS}. Next, we study this dependence in the MR-type and GG-type nematic models introduced above, and show that the results are valid for other similar strain-energy density functions.

\subsection{Stretching of Mooney-Rivlin-type nematic material} 

We first specialise our analysis to the nematic constitutive model combining standard and neoclassical Mooney-Rivlin-type strain-energy densities defined by \eqref{LCE:eq:W:Fn:MR}. We then show that the results obtained in this case are valid for the Gent-Gent-type constitutive relation as well.
 
By substituting $\textbf{F}$ given by \eqref{LCE:F:stretch} in the expression of the strain-energy function, and denoting $w(\lambda,\varepsilon,\theta)=W(\textbf{F},\textbf{n})$, we obtain
\begin{equation}\label{LCE:eq:W:Fn:MR:stretch}
\begin{split}
w(\lambda,\varepsilon,\theta)
&=\frac{\mu^{(1)}_{1}+\mu^{(1)}_{2}}{2}\left(a^{1/3}\lambda^2+a^{-2/3}\varepsilon^2+a^{-1/3}\lambda^{-2}\right)\\
&+\frac{\mu^{(2)}_{1}}{2}a^{1/3}\left\{a^{-1/3}+\lambda^2+\varepsilon^2+a^{1/3}\lambda^{-2}\right.\\
&\left.-\left(1- a^{-1}\right)\left[\left(\lambda^2+\varepsilon^2\right)\sin^2\theta+\varepsilon a^{1/6}\lambda^{-1}\sin\left(2\theta\right)+a^{1/3}\lambda^{-2}\cos^2\theta\right]\right\}\\
&+\frac{\mu^{(2)}_{2}}{2}a^{-1/3}\left\{a^{1/3}+\lambda^{-2}+a^{-1/3}\left(\lambda^2+\varepsilon^2\right)\right.\\
&\left.-\left(1-a\right)\left[\lambda^{-2}\sin^2\theta-\varepsilon a^{-1/6}\lambda^{-1}\sin\left(2\theta\right)+a^{-1/3}\left(\lambda^2+\varepsilon^2\right)\cos^2\theta\right]\right\}.
\end{split}
\end{equation}
Differentiating the above function with respect to $\varepsilon$ and $\theta$, respectively, gives
\begin{equation}\label{LCE:eq:dW:dvarepsilon:MR:stretch}
\begin{split}
\frac{\partial w(\lambda,\varepsilon,\theta)}{\partial\varepsilon}
&=\left(\mu^{(1)}_{1}+\mu^{(1)}_{2}\right)a^{-2/3}\varepsilon\\
&+\frac{\mu^{(2)}_{1}}{2}a^{1/3}\left\{2\varepsilon-\left(1- a^{-1}\right)\left[2\varepsilon\sin^2\theta+a^{1/6}\lambda^{-1}\sin\left(2\theta\right)\right]\right\}\\
&+\frac{\mu^{(2)}_{2}}{2}a^{-1/3}\left\{2\varepsilon a^{-1/3}-\left(1-a\right)\left[2\varepsilon a^{-1/3}\cos^2\theta-a^{-1/6}\lambda^{-1}\sin\left(2\theta\right)\right]\right\},
\end{split}
\end{equation}
\begin{equation}\label{LCE:eq:dW:dtheta:MR:stretch}
\begin{split}
\frac{\partial w(\lambda,\varepsilon,\theta)}{\partial\theta}&=\frac{\mu^{(2)}_{1}}{2}\left(a^{-2/3}-a^{1/3}\right)\left[\left(\lambda^2+\varepsilon^2\right)\sin\left(2\theta\right)+2\varepsilon a^{1/6}\lambda^{-1}\cos\left(2\theta\right)-a^{1/3}\lambda^{-2}\sin\left(2\theta\right)\right]\\
&+\frac{\mu^{(2)}_{2}}{2}\left(a^{2/3}-a^{-1/3}\right)\left[\lambda^{-2}\sin\left(2\theta\right)-2\varepsilon a^{-1/6}\lambda^{-1}\cos\left(2\theta\right)-a^{-1/3}\left(\lambda^2+\varepsilon^2\right)\sin\left(2\theta\right)\right].
\end{split}
\end{equation}
The equilibrium solution minimises the energy, and therefore satisfies
\begin{eqnarray}
&&\frac{\partial w(\lambda,\varepsilon,\theta)}{\partial\varepsilon}=0,\label{LCE:eq:dwdeps}\\
&&\frac{\partial w(\lambda,\varepsilon,\theta)}{\partial\theta}=0.\label{LCE:eq:dwdtheta}
\end{eqnarray}
At $\varepsilon=0$ and $\theta=0$, both the partial derivatives defined by \eqref{LCE:eq:dW:dvarepsilon:MR:stretch}-\eqref{LCE:eq:dW:dtheta:MR:stretch} are equal to zero, i.e.,
\begin{equation}\label{LCE:eq:dwdepsdwdtheta0}
\frac{\partial w}{\partial\varepsilon}(\lambda,0,0)=\frac{\partial w}{\partial\theta}(\lambda,0,0)=0. 
\end{equation}
 Therefore, this trivial solution is always an equilibrium state, and for sufficiently small values of $\varepsilon$ and $\theta$, we have the second order approximation \cite{DeSimone:2009:dST}
\begin{equation}\label{LCE:eq:W:2order:stretch}
w(\lambda,\varepsilon,\theta)\approx w(\lambda,0,0)+\frac{1}{2}\left(\varepsilon^2\frac{\partial^2 w}{\partial\varepsilon^2}(\lambda,0,0)+2\varepsilon\theta\frac{\partial^2 w}{\partial\varepsilon\partial\theta}(\lambda,0,0)+\theta^2\frac{\partial^2 w}{\partial\theta^2}(\lambda,0,0)\right),
\end{equation}
where
\begin{eqnarray}
&&\frac{\partial^2 w}{\partial\varepsilon^2}(\lambda,0,0)=\mu^{(1)}a^{-2/3}+\mu^{(2)}a^{1/3},\\
&&\frac{\partial^2 w}{\partial\varepsilon\partial\theta}(\lambda,0,0)=\frac{\mu^{(2)}}{\lambda}\left(a^{-1/2}-a^{1/2}\right),\\
&&\frac{\partial^2 w}{\partial\theta^2}(\lambda,0,0)=\mu^{(2)}\left(a^{-2/3}-a^{1/3}\right)\left(\lambda^2-a^{1/3}\lambda^{-2}\right),\label{LCE:eq:d2w:d2theta:stretch}
\end{eqnarray}
with $\mu^{(1)}=\mu^{(1)}_{1}+\mu^{(1)}_{2}$ and  $\mu^{(2)}=\mu^{(2)}_{1}+\mu^{(2)}_{2}$. First, we find the equilibrium value $\theta_{0}$ for $\theta$ as a function of $\varepsilon$ by solving equation \eqref{LCE:eq:dwdtheta}, which, after the approximation \eqref{LCE:eq:W:2order:stretch}, takes the form
\begin{equation}\label{LCE:eq:thetaepsilon}
\varepsilon\frac{\partial^2 w}{\partial\varepsilon\partial\theta}(\lambda,0,0)+\theta\frac{\partial^2 w}{\partial\theta^2}(\lambda,0,0)=0,
\end{equation}
and implies
\begin{equation}\label{LCE:eq:theta0epsilon}
\theta_{0}(\varepsilon)=-\varepsilon\frac{\partial^2 w}{\partial\varepsilon\partial\theta}(\lambda,0,0)/\frac{\partial^2 w}{\partial\theta^2}(\lambda,0,0).
\end{equation}
Next, substituting $\theta=\theta_{0}(\varepsilon)$ in \eqref{LCE:eq:W:2order:stretch} gives the following function of $\varepsilon$,
\begin{equation}\label{LCE:eq:stability}
w(\lambda,\varepsilon,\theta_{0}(\varepsilon))-w(\lambda,0,0)\approx\frac{\varepsilon^2}{2}\left[\frac{\partial^2 w}{\partial\varepsilon^2}(\lambda,0,0)-\left(\frac{\partial^2 w}{\partial\varepsilon\partial\theta}(\lambda,0,0)\right)^2/\frac{\partial^2 w}{\partial\theta^2}(\lambda,0,0)\right].
\end{equation}
Depending on whether the expression on the right-hand side in \eqref{LCE:eq:stability} is positive, zero, or negative, the equilibrium state with $\varepsilon=0$ and $\theta=0$ is stable, neutrally stable, or unstable \cite{DeSimone:2009:dST}. Clearly, if $\mu^{(2)}=0$ (purely elastic case) and $\mu^{(1)}>0$, then the equilibrium state is stable. When $\mu^{(2)}>0$, assuming $a>1$, our calculations show that the equilibrium state is unstable for
\begin{equation}\label{LCE:eq:bounds1}
a^{1/12}\left(\frac{\mu^{(1)}/\mu^{(2)}+1}{\mu^{(1)}/\mu^{(2)}+a}\right)^{1/4}<\lambda<a^{1/12}.
\end{equation}
Note that the upper bound in \eqref{LCE:eq:bounds1} is independent of $\mu^{(1)}$ and $\mu^{(2)}$. This is due to the fact that this bound is obtained from the condition $\frac{\partial^2 w}{\partial\theta^2}(\lambda,0,0)>0$, which by \eqref{LCE:eq:d2w:d2theta:stretch}, only involves the neoclassical component of the strain-energy function. The lower bound in \eqref{LCE:eq:bounds1} satisfies
\begin{equation}\label{LCE:eq:lowbound1}
a^{-1/6}\leq a^{1/12}\left(\frac{\mu^{(1)}/\mu^{(2)}+1}{\mu^{(1)}/\mu^{(2)}+a}\right)^{1/4},
\end{equation}
where the equality holds if $\mu^{(1)}=0$ (neoclassical model), in which case the trivial solution is unstable for $\lambda\in\left(a^{-1/6}, a^{1/12}\right)$, as is well known \cite{Conti:2002:CdSD,DeSimone:2009:dST}. 

There is also another equilibrium state, with  $\varepsilon=0$ and $\theta=\pi/2$, where the nematic director is fully rotated so that it aligns uniformly with the direction of macroscopic extension. In this case, by analogous calculations (or by symmetry arguments), we obtain that this state is unstable for
\begin{equation}\label{LCE:eq:bounds2}
a^{1/12}<\lambda<a^{1/12}\left(\frac{\mu^{(1)}/\mu^{(2)}+a}{\mu^{(1)}/\mu^{(2)}+1}\right)^{1/4}.
\end{equation}
Note that the upper bound in \eqref{LCE:eq:bounds2} satisfies
\begin{equation}\label{LCE:eq:highbound2}
a^{1/12}\left(\frac{\mu^{(1)}/\mu^{(2)}+a}{\mu^{(1)}/\mu^{(2)}+1}\right)^{1/4}<a^{1/3}.
\end{equation}

\begin{figure}[htbp]
	\begin{center}
		\includegraphics[width=0.6\textwidth]{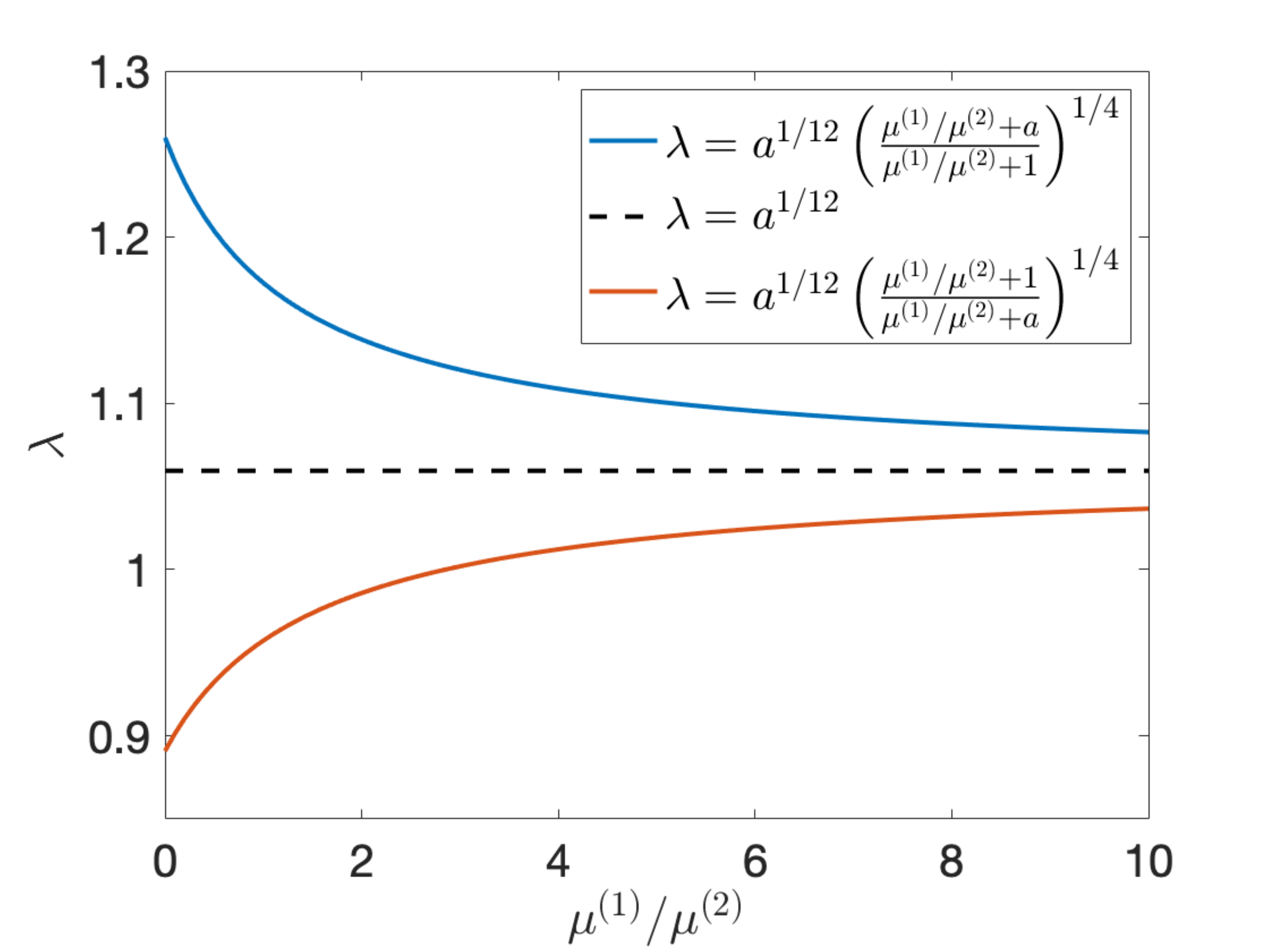}
		\caption{The lower and upper bounds on $\lambda$, given by \eqref{LCE:eq:bounds1} and \eqref{LCE:eq:bounds2}, respectively, as functions of the parameter ratio $\mu^{(1)}/\mu^{(2)}$ when $a=2$. For $\mu^{(1)}/\mu^{(2)}\to\infty$, the model approaches a purely elastic form where the homogeneous deformation is always stable. For $\mu^{(1)}/\mu^{(2)}=0$, the bounds are the same as for the neoclassical form.}\label{LCE:fig:bounds}
	\end{center}
\end{figure}

For illustration, setting $a=2$, the lower and upper bounds given by \eqref{LCE:eq:bounds1}  and \eqref{LCE:eq:bounds2}, respectively, are represented as functions of the parameter ratio $\mu^{(1)}/\mu^{(2)}$ in Figure~\ref{LCE:fig:bounds}. These bounds behave similarly to those shown in Figure~2 of \cite{Fried:2006:FS}.

\begin{figure}[htbp]
	\begin{center}
		\includegraphics[width=0.49\textwidth]{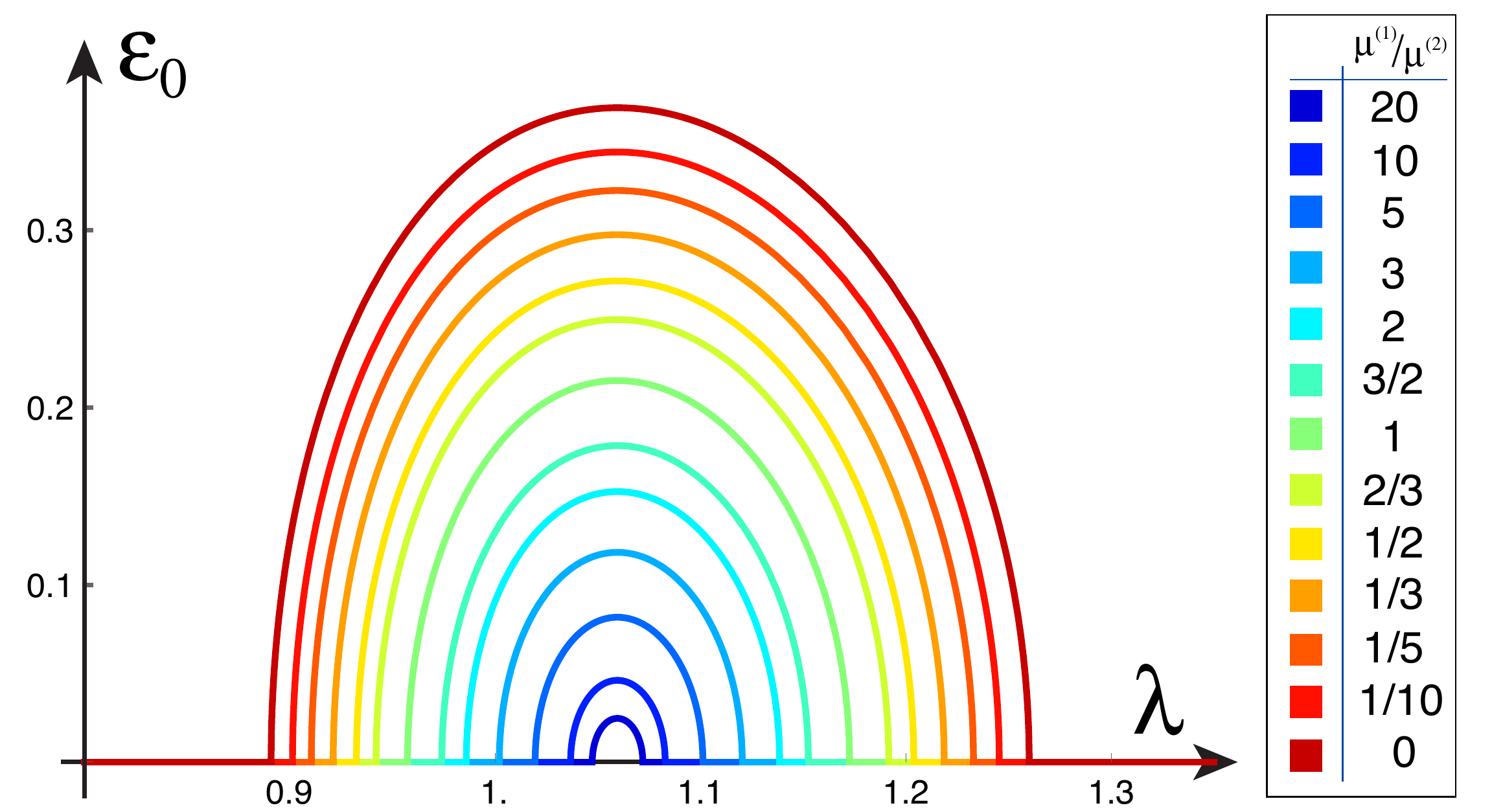}
		\includegraphics[width=0.49\textwidth]{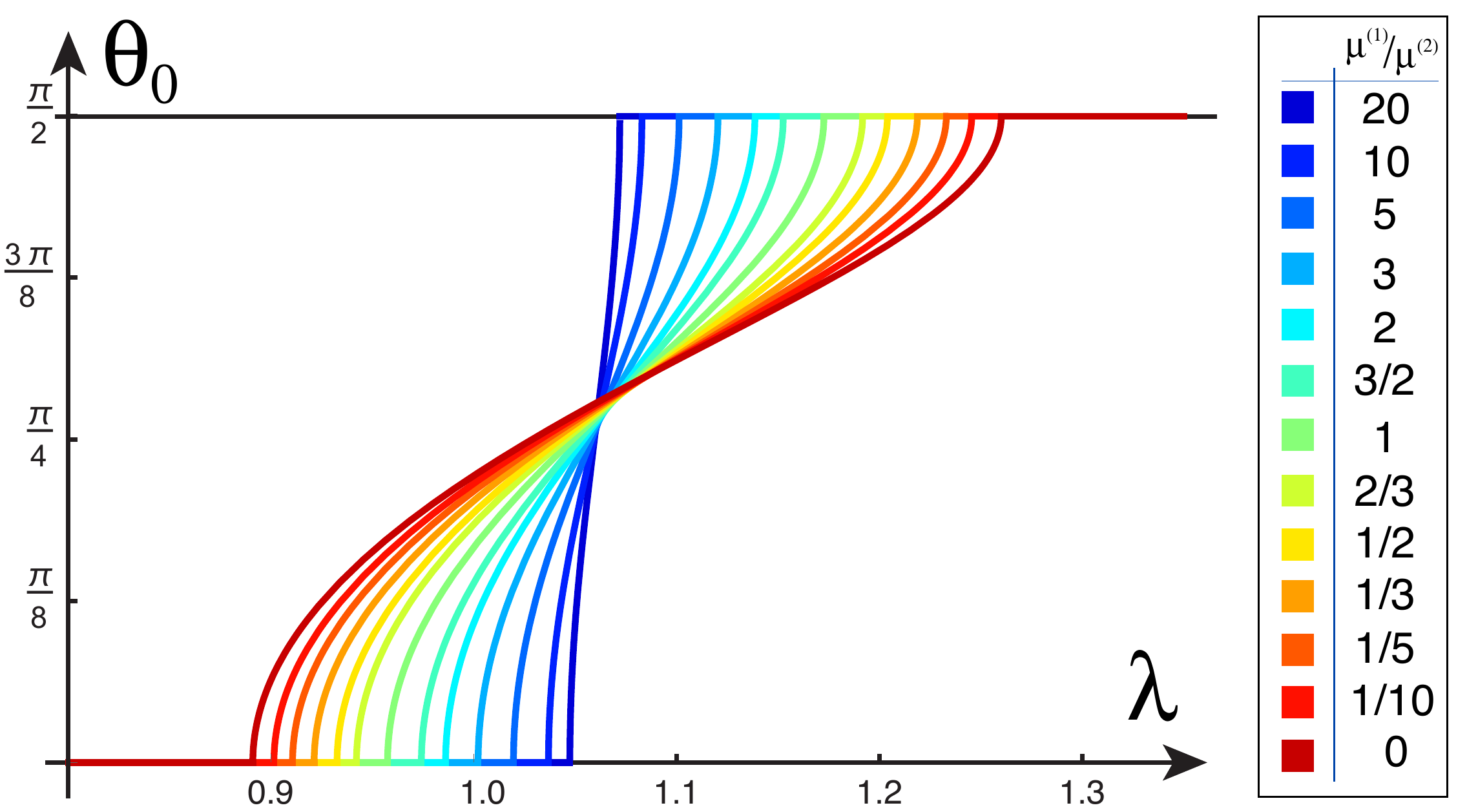}
		\caption{The positive values of $\varepsilon_{0}$ and $\theta_{0}$, given by equations \eqref{LCE:eq:eps0}-\eqref{LCE:eq:theta0}, as functions of $\lambda$, for different parameter ratios $\mu^{(1)}/\mu^{(2)}$ when $a=2$.}\label{LCE:fig:solution}
	\end{center}
\end{figure}

It remains to find the inhomogeneous solution in the case when $\lambda$ satisfies either \eqref{LCE:eq:bounds1}  or \eqref{LCE:eq:bounds2}. For the equilibrium angle $\theta_{0}(\varepsilon)$, given by \eqref{LCE:eq:theta0epsilon}, equation \eqref{LCE:eq:dwdtheta} is satisfied. To find the equilibrium value $\varepsilon_{0}$ of $\varepsilon$ as a function of $\lambda$ satisfying \eqref{LCE:eq:bounds1} and \eqref{LCE:eq:bounds2}, we solve the simultaneous equations \eqref{LCE:eq:dwdeps}-\eqref{LCE:eq:dwdtheta}, and obtain
\begin{eqnarray}
&&\varepsilon_{0}=\pm\frac{\lambda(a-1)\sin(2\theta_{0})}{2\sqrt{\left(\mu^{(1)}/\mu^{(2)}+1\right)\left(\mu^{(1)}/\mu^{(2)}+a\right)}},\label{LCE:eq:eps0}\\
&&\theta_{0}=\pm\arccos\sqrt{\frac{a^{1/6}\sqrt{\left(\mu^{(1)}/\mu^{(2)}+1\right)\left(\mu^{(1)}/\mu^{(2)}+a\right)}-\lambda^2\left(\mu^{(1)}/\mu^{(2)}+1\right)}{\lambda^2(a-1)}}.\label{LCE:eq:theta0}
\end{eqnarray}

\begin{figure}[htbp]
	\begin{center}
		\includegraphics[width=0.6\textwidth]{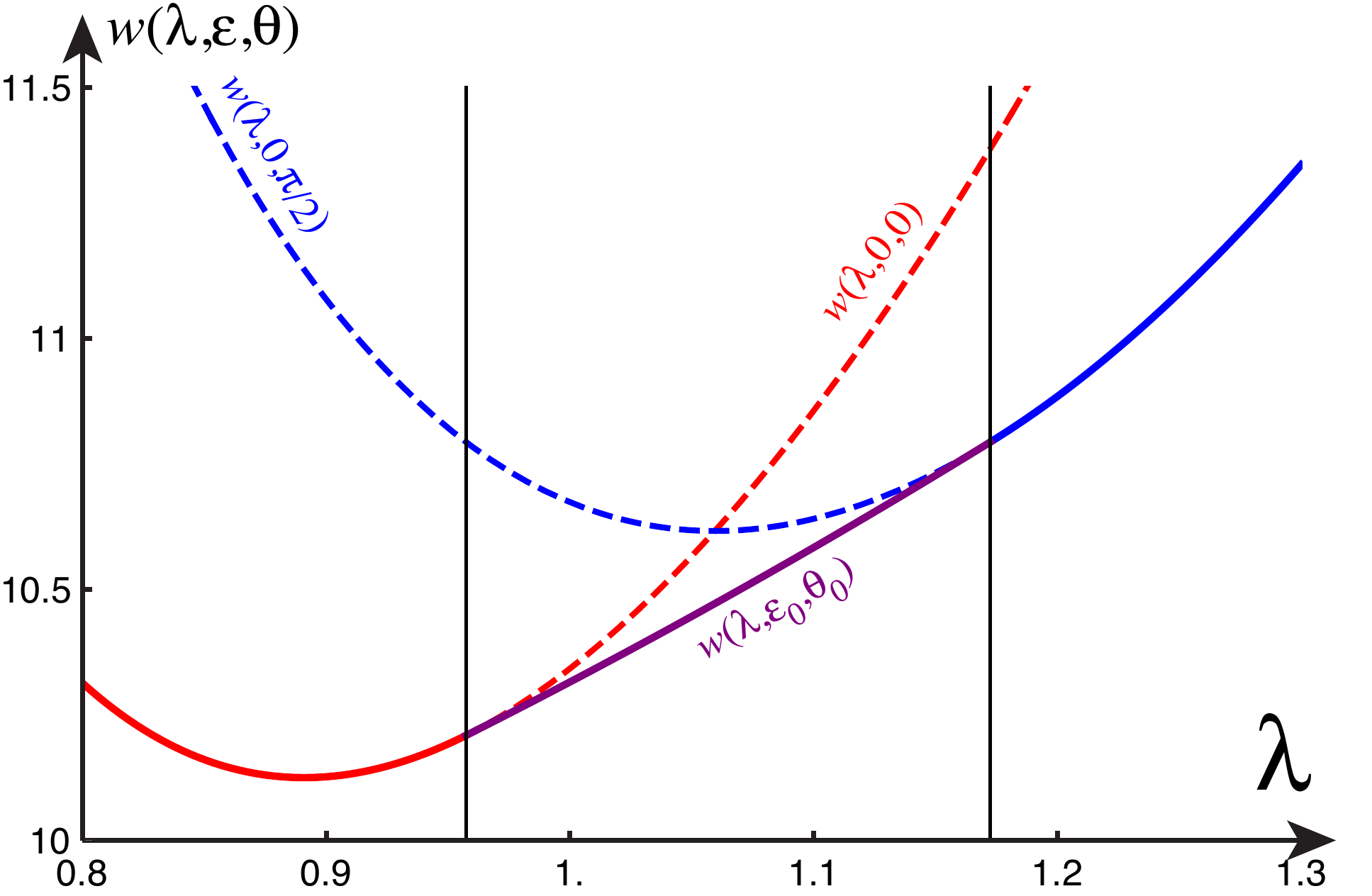}
		\caption{The strain-energy function $w(\lambda,\varepsilon,\theta)$, defined by equation \eqref{LCE:eq:W:Fn:MR:stretch}, for $(\varepsilon,\theta)=(0,0)$, $(\varepsilon,\theta)=(\varepsilon_0,\theta_0)$, with $\varepsilon_{0}$ and $\theta_{0}$ given by \eqref{LCE:eq:eps0}-\eqref{LCE:eq:theta0}, and $(\varepsilon,\theta)=(0,\pi/2)$, when $a=2$ and $\mu^{(1)}=\mu^{(2)}=4.05$, with $\mu^{(2)}_{1}=2.05$ and $\mu^{(2)}_{2}=2$. The two vertical lines correspond to the lower and upper bounds on $\lambda$, given by equations \eqref{LCE:eq:bounds1}  and \eqref{LCE:eq:bounds2}, respectively. Between these bounds, the second solution, with $(\varepsilon,\theta)=(\varepsilon_0,\theta_0)$, minimises the energy.}\label{LCE:fig:energy}
	\end{center}
\end{figure}

For example, setting $a=2$ and taking different parameter ratios $\mu^{(1)}/\mu^{(2)}$, the positive values of $\varepsilon_{0}$ and $\theta_{0}$, given by \eqref{LCE:eq:eps0}-\eqref{LCE:eq:theta0}, are illustrated in Figure~\ref{LCE:fig:solution} . These values are comparable to those shown the Figures~5 and 6 of \cite{Fried:2006:FS} (see also Figure~5 of \cite{Verwey:1996:VWT}). In particular, for $a=2$ and $\mu^{(1)}=\mu^{(2)}=4.05$, such that $\mu^{(2)}_{1}=2.05$ and $\mu^{(2)}_{2}=2$, the strain-energy function $w(\lambda,\varepsilon,\theta)$, defined by \eqref{LCE:eq:W:Fn:MR:stretch}, is represented in Figure~\ref{LCE:fig:energy} when $(\varepsilon,\theta)\in\left\{(0,0),(\varepsilon_0,\theta_0),(0,\pi/2)\right\}$. As seen from this figure, for $\lambda$ with values between the lower and upper bounds given by \eqref{LCE:eq:bounds1} and  \eqref{LCE:eq:bounds2}, respectively, the minimum energy is attained for $(\varepsilon,\theta)=(\varepsilon_0,\theta_0)$. These results are (qualitatively) similar with those shown in Figure~2 of \cite{Fried:2006:FS} (see also Figure~3 of \cite{Verwey:1996:VWT}, or Figure~7.12 of \cite[p.~180]{Warner:2007:WT}).

\subsection{Stretching of Gent-Gent-type nematic material} 

Next, we consider the nematic model that combines classical and neoclassical Gent-Gent-type strain-energy densities, given by \eqref{LCE:eq:W:Fn:GG} (the idea of considering different constitutive relations was suggested also in the discussion section of \cite{Fried:2006:FS}). Substituting $\textbf{F}$ defined by \eqref{LCE:F:stretch} in the expression of the strain-energy function, and denoting $w(\lambda,\varepsilon,\theta)=W(\textbf{F},\textbf{n})$, gives
\begin{equation}\label{LCE:eq:W:Fn:GG:stretch}
\begin{split}
w(\lambda,\varepsilon,\theta)
&=-\frac{\mu^{(1)}_{1}}{2\beta^{(1)}}\ln\left[1-\beta^{(1)}\left(a^{1/3}\lambda^2+\varepsilon^2+a^{-1/3}\lambda^{-2}-2\right)\right]\\
&+\frac{3\mu^{(1)}_{2}}{2}\ln\left(1+a^{-1/3}\lambda^{-2}+a^{1/3}\lambda^2+\varepsilon^2\right)\\
&-\frac{3\mu^{(1)}_{2}}{2}\ln3\\
&-\frac{\mu^{(2)}_{1}}{2\beta^{(2)}}\ln\left\{1-\beta^{(2)} a^{1/3}\left\{a^{-1/3}+\lambda^2+\varepsilon^2+a^{1/3}\lambda^{-2}\right.\right.\\
&\left.\left.-\left(1- a^{-1}\right)\left[\left(\lambda^2+\varepsilon^2\right)\sin^2\theta+\varepsilon a^{1/6}\lambda^{-1}\sin\left(2\theta\right)+a^{1/3}\lambda^{-2}\cos^2\theta\right]\right\}+3\beta^{(2)}\right\}\\
&+\frac{3\mu^{(2)}_{2}}{2}\ln\left\{a^{-1/3}\left\{a^{1/3}+\lambda^{-2}+a^{-1/3}\left(\lambda^2+\varepsilon^2\right)\right.\right.\\
&\left.\left.-\left(1-a\right)\left[\lambda^{-2}\sin^2\theta-\varepsilon a^{-1/6}\lambda^{-1}\sin\left(2\theta\right)+a^{-1/3}\left(\lambda^2+\varepsilon^2\right)\cos^2\theta\right]\right\}\right\}\\
&-\frac{3\mu^{(2)}_{2}}{2}\ln3.
\end{split}
\end{equation}

For this material also, an equilibrium state is obtained when $\varepsilon=0$ and $\theta=0$. Then, the function $w(\lambda,\varepsilon,\theta)$ takes the approximate form \eqref{LCE:eq:W:2order:stretch}, and depending on whether the associated expression given by \eqref{LCE:eq:stability} is positive, zero, or negative, this equilibrium state is stable, neutrally stable, or unstable. Then, similarly to the case of the MR-type nematic material, the equilibrium state is always stable when $\mu^{(2)}=\mu^{(2)}_{1}+\mu^{(2)}_{2}=0$ and $\mu^{(1)}=\mu^{(1)}_{1}+\mu^{(1)}_{2}>0$. When $\mu^{(2)}>0$, it can be verified that, under the assumption (A3) that the BE inequalities \eqref{LCE:eq:Wnc:BE} hold, for $a>1$, the equilibrium solution is unstable for $\lambda$ satisfying \eqref{LCE:eq:bounds1}. If $\mu^{(1)}=0$, then the equilibrium state is unstable for $\lambda\in\left(a^{-1/6}, a^{1/12}\right)$ \cite{Conti:2002:CdSD,DeSimone:2009:dST}. The analysis of the equilibrium state at $\varepsilon=0$ and $\theta=\pi/2$ is analogous. Therefore, the results obtained for the MR-type model are also valid for the GG-type model.

\subsection{Stretching of nematic solids with stochastic parameters} 

We now examine the shear striping problem when the model parameters are cast as random variables following spatially-independent probability distributions. In this case, one can regard the stretched sample of stochastic nematic elastomer as an ensemble (or population) of samples with the same geometry, such that each sample is made from a homogeneous nematic material with the model parameters not known with certainty, but drawn from known probability distributions. Then, for each individual sample, the finite strain theory for homogeneous materials applies. The question is: \emph{what is the probability that shear striping occurs under a given stretch?} In what follows, we answer this question by showing the effects of fluctuations in the shear modulus, $\mu$, and the nematic parameter, $a$, separately, i.e., we first set $\mu$ as a random variable while $a$ is constant, then keep $\mu$ constant and let $a$ fluctuate. The problem where both these parameters are stochastic requires a detailed analysis of their statistical relation (see, e.g., \cite{Staber:2018:SG}), and will be treated in detail somewhere else.

\subsubsection{Stochastic shear modulus and deterministic nematic stretch parameter}
First, we assume that the shear modulus $\mu=\mu^{(1)}+\mu^{(2)}=\mu^{(1)}_{1}+\mu^{(1)}_{2}+\mu^{(2)}_{1}+\mu^{(2)}_{2}$ is a random variable, such that $\mu^{(1)}=\mu^{(1)}_{1}+\mu^{(1)}_{2}>0$ and $\mu^{(2)}=\mu^{(2)}_{1}+\mu^{(2)}_{2}>0$, while $a$ is deterministic, such that $a>1$. For purely elastic models, probability distributions for the individual model coefficients contributing to the shear parameter $\mu$ are discussed in \cite{Mihai:2019a:MDWG,Mihai:2019c:MDWG,Mihai:2018:MWG,Mihai:2019a:MWG}. Here, we only require the probability density function of $\mu$, given by \eqref{eq:mu:gamma}.

Substituting $\mu^{(2)}=\mu-\mu^{(1)}$ in the inequality involving the lower bound on $\lambda$ in \eqref{LCE:eq:bounds1} implies
\begin{equation}
\frac{\mu}{\mu^{(1)}}>\left(1-\frac{1}{a}\right)\frac{\lambda^{4}}{\lambda^{4}-a^{-2/3}}.
\end{equation}
Hence, given $a>1$ and $\lambda>0$, for any $\mu^{(1)}>0$, the probability that the equilibrium state at $\varepsilon=0$ and $\theta=0$ is unstable, such that shear striping develops, is
\begin{equation}\label{LCE:eq:mu1:P1:left}
P_{1}(\mu^{(1)})=1-\int_{0}^{\mu^{(1)}\left(1-\frac{1}{a}\right)\frac{\lambda^{4}}{\lambda^{4}-a^{-2/3}}}g(u;\rho_{1},\rho_{2})\text{d}u,
\end{equation}
while the probability of stable equilibrium at $\varepsilon=0$ and $\theta=0$ is 
\begin{equation}\label{LCE:eq:mu1:P2:left}
P_{2}(\mu^{(1)})=1-P_{1}(\mu^{(1)})=\int_{0}^{\mu^{(1)}\left(1-\frac{1}{a}\right)\frac{\lambda^{4}}{\lambda^{4}-a^{-2/3}}}g(u;\rho_{1},\rho_{2})\text{d}u,
\end{equation}
where $g(u;\rho_{1},\rho_{2})$ is defined by \eqref{eq:mu:gamma}.

\begin{figure}[htbp]
	\begin{center}
		\includegraphics[width=0.99\textwidth]{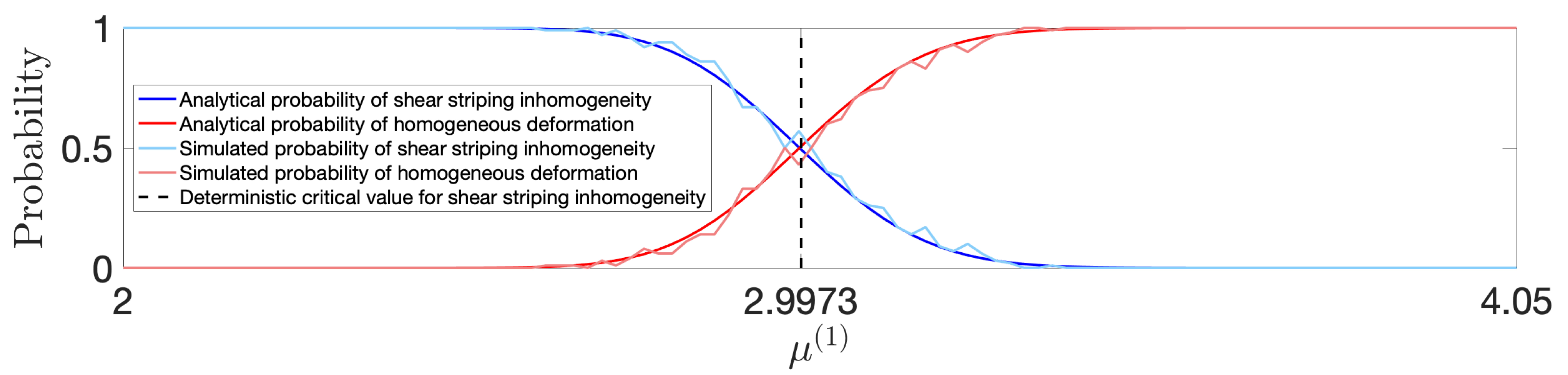}
		\caption{Probability distributions of whether shear striping can occur or not for a  monodomain nematic elastomer where $a=2$, $\lambda=1$, and the shear modulus, $\mu$, is drawn from the Gamma distribution with $\rho_{1}=405$, $\rho_{2}=0.01$. Dark coloured lines represent analytically derived solutions, given by equations  \eqref{LCE:eq:mu1:P1:left}-\eqref{LCE:eq:mu1:P2:left}, whereas the lighter versions represent stochastically generated data. The vertical line at the critical value $\mu^{(1)}=2.9973$ corresponds to the deterministic solution based only on the mean value $\underline{\mu}=\rho_{1}\rho_{2}=4.05$. The probabilities were calculated from the average of 100 stochastic simulations.}\label{LCE:fig:stripes-intpdfs}
	\end{center}
	\begin{center}
		\includegraphics[width=0.49\textwidth]{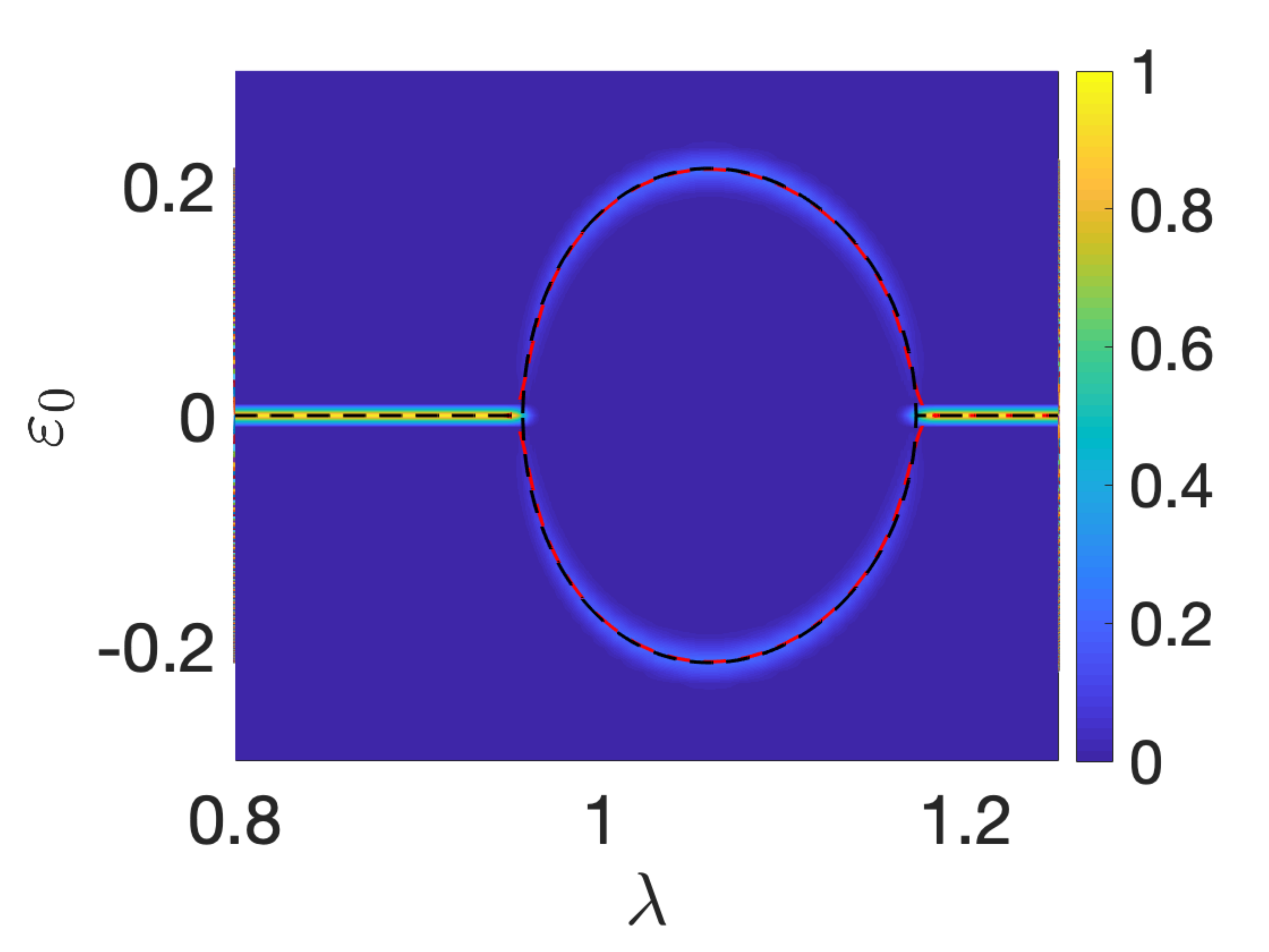}
		\includegraphics[width=0.49\textwidth]{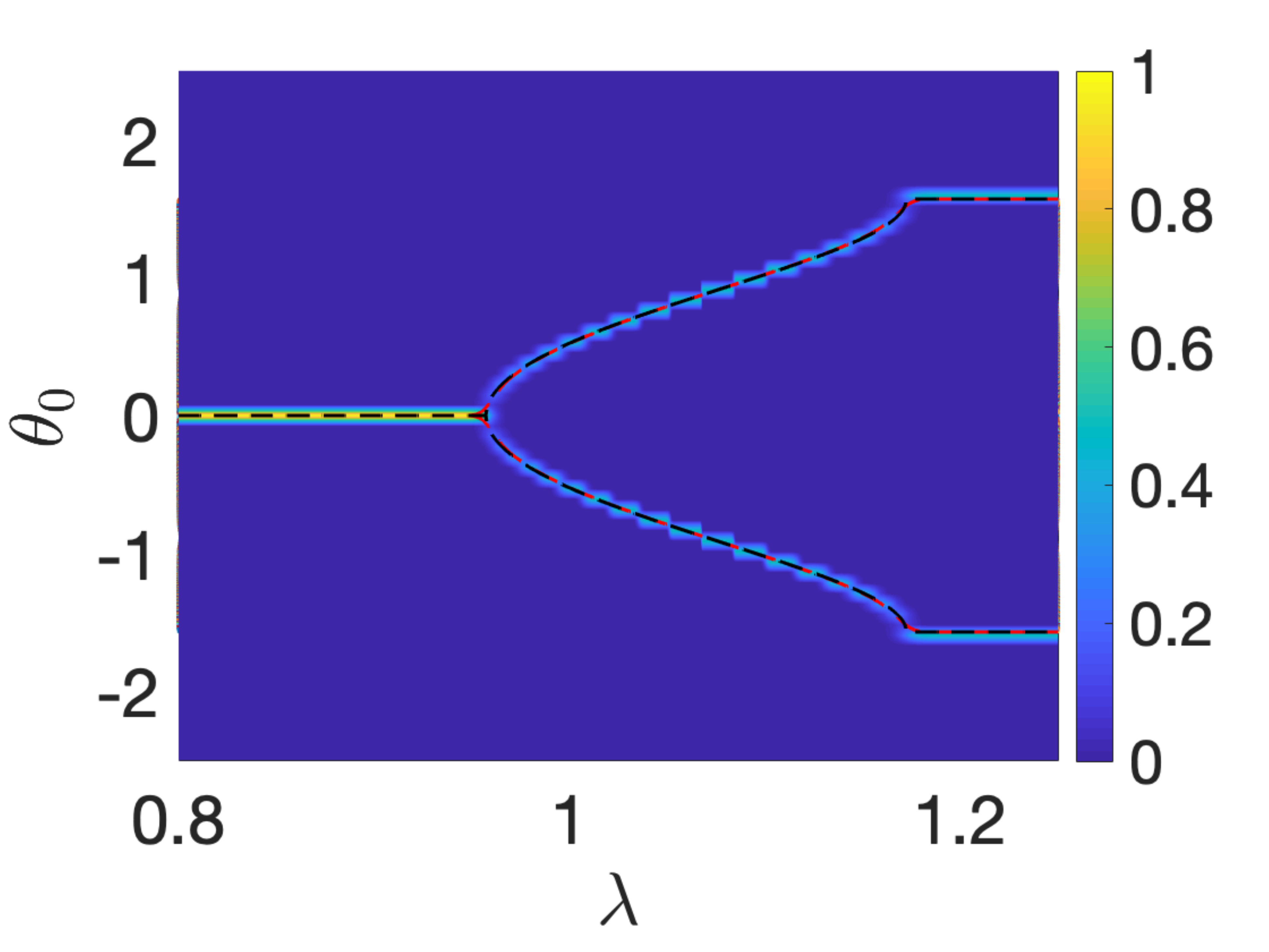}
		\caption{The stochastic shear parameter $\varepsilon_{0}$ and director angle $\theta_{0}$, given by \eqref{LCE:eq:eps0}-\eqref{LCE:eq:theta0}, when $a=2$ and $\mu=\mu^{(1)}+\mu^{(2)}$ follows a Gamma probability distribution with hyperparameters $\rho_{1}=405$, $\rho_{2}=0.01$, such that $\mu^{(1)}=4.05/2$ is deterministic. Each distribution was calculated from the average of $1000$ stochastic simulations. The dashed black lines correspond to the deterministic solutions based only on the mean value $\underline{\mu}=\rho_{1}\rho_{2}=4.05$, whereas the red versions show the arithmetic mean value solutions computed numerically.}\label{LCE:fig:mustoch}
	\end{center}
\end{figure}

For example, setting $a=2$ and $\lambda=1$, the probability distributions given by equations \eqref{LCE:eq:mu1:P1:left}-\eqref{LCE:eq:mu1:P2:left} are illustrated both analytically and numerically in Figure~\ref{LCE:fig:stripes-intpdfs} (blue lines for $P_{1}$ and red lines for $P_{2}$). For the numerical approximation, the interval $(2,\underline{\mu})$ for $\mu^{(1)}$ was discretised into $100$ points, then for each value of $\mu^{(1)}$, $100$ random values of $\mu$ were numerically generated from a specified Gamma distribution, with hyperparameters $\rho_{1}=405$, $\rho_{2}=0.01$, and compared with the inequalities defining the two intervals for values of $\mu^{(1)}$. Note that, although only 100 random numbers were generated, the numerical approximation captures well the analytical solution. Our simulations were run in Matlab 2019a, where we made specific use of inbuilt functions for random number generation. 

In the deterministic case based on the mean value of the shear modulus, $\underline{\mu}=\rho_{1}\rho_{2}=4.05$, the critical value of $\mu^{(1)}=\underline{\mu}/1.3512=2.9973$ strictly divides the regions where striping inhomogeneity can occur or not. For the stochastic problem, to increase the probability of homogeneous deformations ($P_{2}\approx 1$), one must consider sufficiently large values of $\mu^{(1)}$, whereas shear striping is certain ($P_{1}\approx 1$) only if the model reduces to the neoclassical one (i.e., when $\mu^{(1)}=0$). However, while $\mu^{(1)}>0$, the inherent variability in the probabilistic system means that there will always be competition between the homogeneous and inhomogeneous deformations. 

In Figure~\ref{LCE:fig:mustoch}, the stochastic shear parameter $\varepsilon_{0}$ and director angle $\theta_{0}$, given by \eqref{LCE:eq:eps0}-\eqref{LCE:eq:theta0}, are illustrated when $a=2$ and $\mu=\mu^{(1)}+\mu^{(2)}$ is drawn from a Gamma probability distribution with shape and scale parameters $\rho_{1}=405$ and $\rho_{2}=0.01$, respectively, such that $\mu^{(1)}=4.05/2$ is a deterministic constant. To compare directly our stochastic results with the deterministic ones, we sampled from a distribution where the shear modulus was set to have the mean value corresponding to the deterministic system. Due to the linear form in which this modulus is computed from the model coefficients, the mean value solutions are guaranteed to converge to the deterministic solutions.

\subsubsection{Stochastic nematic stretch parameter and deterministic shear modulus}

Next, we assume that the value of the nematic parameter, $a$, `fluctuates', while the nematic director remains uniformly oriented \cite{Verwey:1995:VW,Verwey:1996:VWT} (see also the discussion in \cite{Fried:2006:FS}). Given that $a$ takes finite positive values, it is reasonable to require that assumption (A4) also holds for this parameter, and so we can modify this assumption by replacing $\mu$ with $a$. Then, under constraints similar to \eqref{eq:Emu1}, by the maximum entropy principle, the random nematic parameter $a$ follows a Gamma probability distribution. This is a prior probability distribution that can be made more precise if experimental data are provided. To study the independent effect of fluctuating $a$, we take the coefficients $\mu^{(1)}$ and $\mu^{(2)}$ to be deterministic constants. Hence $\mu=\mu^{(1)}+\mu^{(2)}$ is deterministic.

For example, setting $\lambda=1$, the inequality involving the lower bound on $\lambda$ in \eqref{LCE:eq:bounds1} is equivalent to
\begin{equation}
a>a_c=\left[\sqrt{\left(\frac{\mu}{\mu^{(1)}}- 1\right)\left(\frac{\mu}{\mu^{(1)}} + 3\right)} - \frac{\mu}{\mu^{(1)}} + 1\right]^3/\left[8\left(\frac{\mu}{\mu^{(1)}} - 1\right)^3\right].
\end{equation}
Hence, for any $\mu/\mu^{(1)}>1$ the probability that the equilibrium state at $\varepsilon=0$ and $\theta=0$ is unstable, such that shear stripes form, is
\begin{equation}\label{LCE:eq:a:P1:left}
P_{1}\left(\frac{\mu}{\mu^{(1)}}\right)=1-\int_{0}^{a_c} g(u;\rho_{1},\rho_{2})\text{d}u,
\end{equation}
while the probability of stable equilibrium at $\varepsilon=0$ and $\theta=0$ is 
\begin{equation}\label{LCE:eq:a:P2:left}
P_{2}\left(\frac{\mu}{\mu^{(1)}}\right)=1-P_{1}\left(\frac{\mu}{\mu^{(1)}}\right)=\int_{0}^{a_c} g(u;\rho_{1},\rho_{2})\text{d}u,
\end{equation}
where $g(u;\rho_{1},\rho_{2})$ now represents the probability density function for the Gamma-distributed $a$.

\begin{figure}[htbp]
	\begin{center}
		\includegraphics[width=0.99\textwidth]{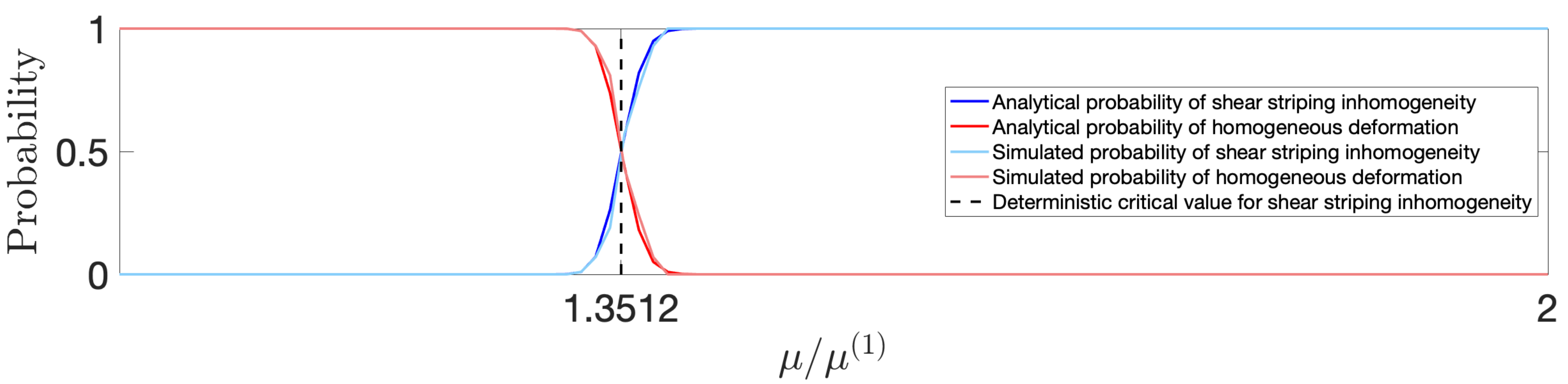}
		\caption{Probability distributions of whether shear striping can occur or not for a  monodomain nematic elastomer where $\lambda=1$ and the nematic stretch parameter, $a$, is drawn from the Gamma distribution with $\rho_{1}=200$, $\rho_{2}=0.01$. Dark coloured lines represent analytically derived solutions, given by equations  \eqref{LCE:eq:a:P1:left}-\eqref{LCE:eq:a:P2:left}, while the lighter colours represent stochastically generated data. The vertical line at the critical value $\mu/\mu^{(1)}=1.3512$ corresponds to the deterministic solution based only on the mean value $\underline{a}=\rho_{1}\rho_{2}=2$. The probabilities were calculated from the average of 100 stochastic simulations.}\label{LCE:fig:stripes-aintpdfs}
	\end{center}
	\begin{center}
		\includegraphics[width=0.49\textwidth]{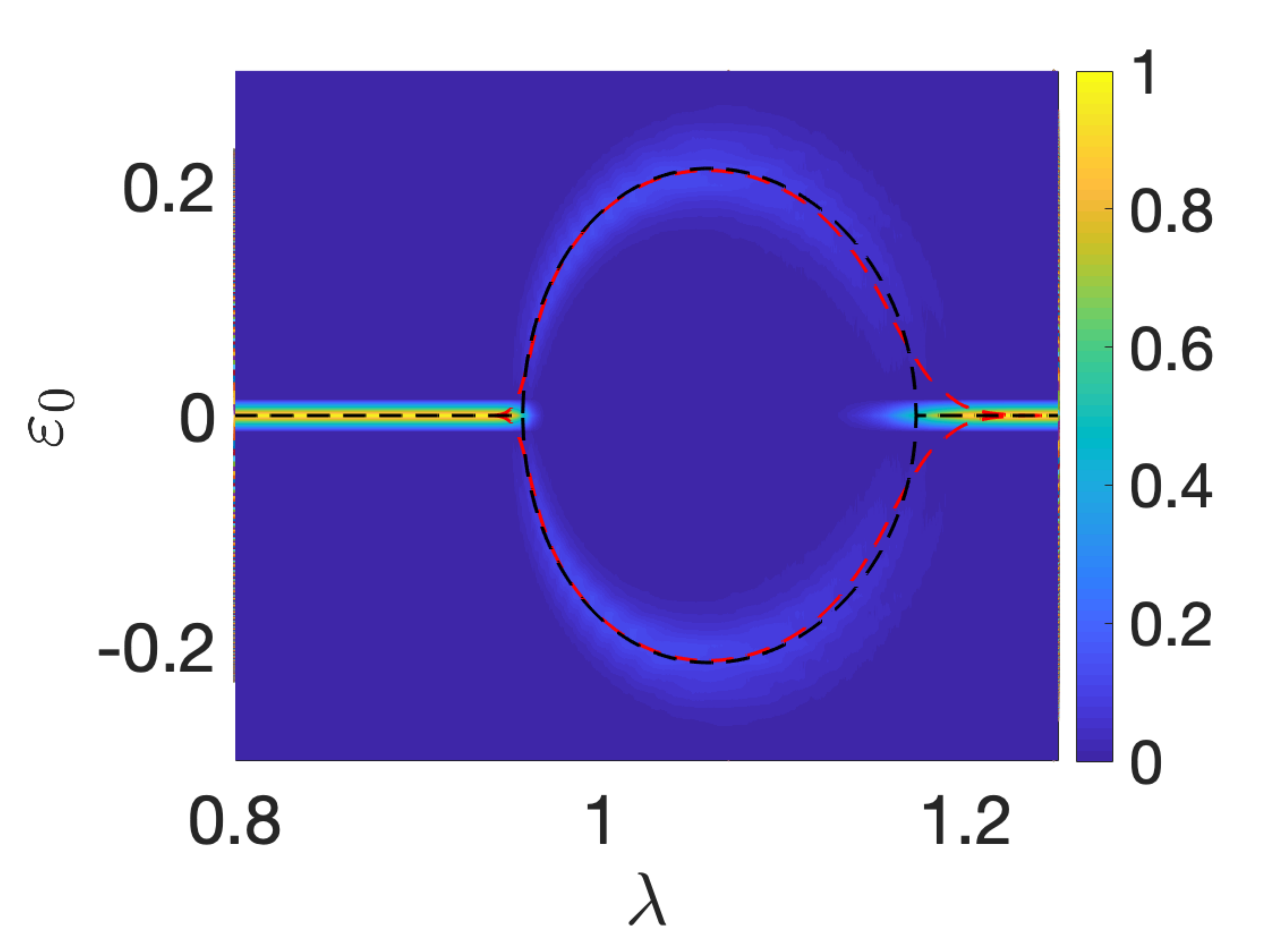}
		\includegraphics[width=0.49\textwidth]{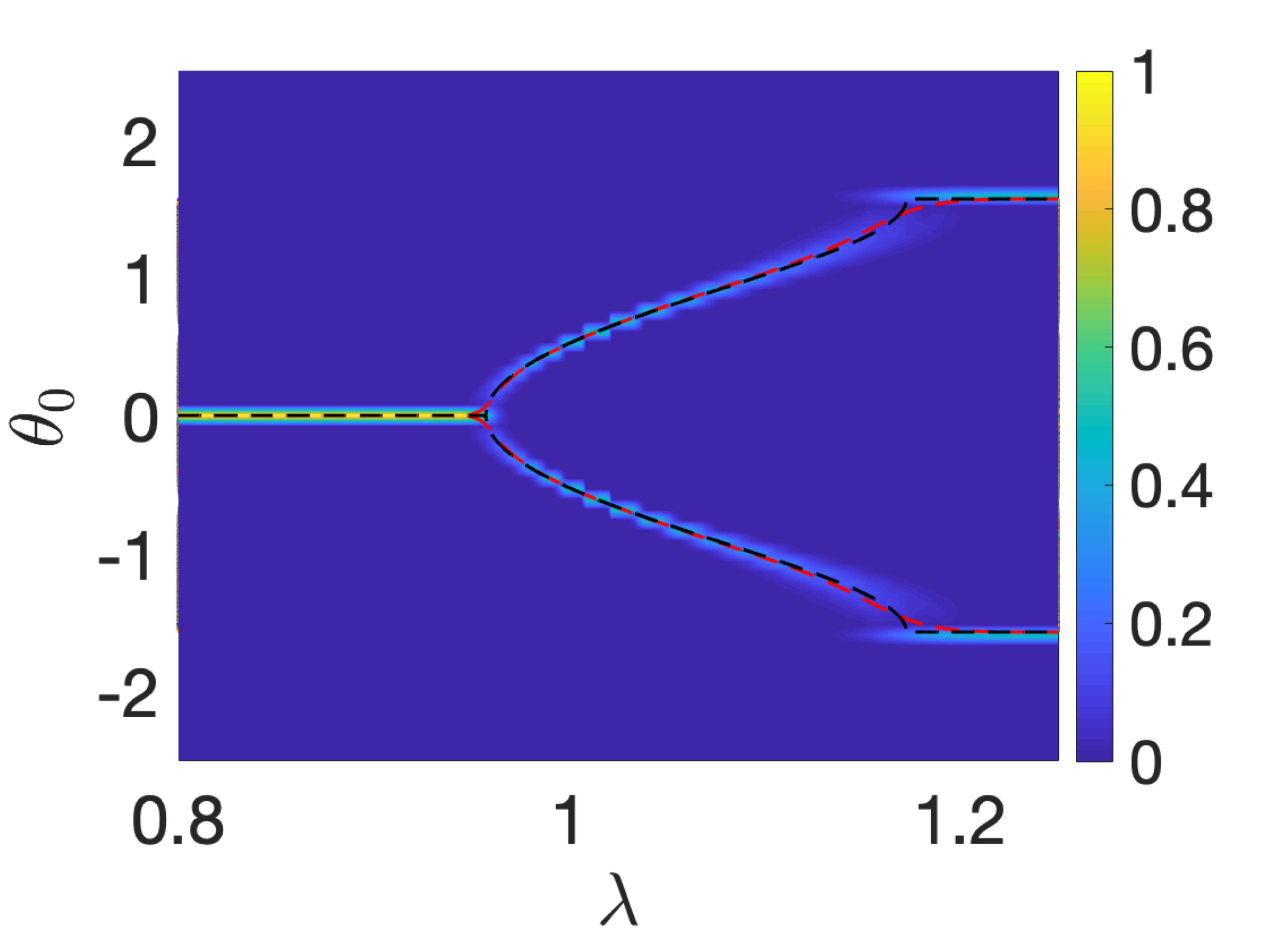}
		\caption{The stochastic shear parameter $\varepsilon_{0}$ and director angle $\theta_{0}$, given by \eqref{LCE:eq:eps0}-\eqref{LCE:eq:theta0}, when $a$ follows a Gamma probability distribution with hyperparameters $\rho_{1}=200$, $\rho_{2}=0.01$, and $\mu=\mu^{(1)}+\mu^{(2)}=4.05$ is deterministic, with $\mu^{(1)}=\mu^{(2)}=4.05/2$. Each distribution was calculated from the average of $1000$ stochastic simulations. The dashed black lines correspond to the deterministic solutions based only on the mean value $\underline{a}=\rho_{1}\rho_{2}=2$, whereas the red versions show the arithmetic mean value solutions computed numerically.}\label{LCE:fig:astoch}
	\end{center}
\end{figure}

The probability distributions given by equations \eqref{LCE:eq:a:P1:left}-\eqref{LCE:eq:mu1:P2:left} are illustrated in Figure~\ref{LCE:fig:stripes-aintpdfs} (blue lines for $P_{1}$ and red lines for $P_{2}$). For the numerical realisations in these plots, the interval $(1,2)$ $\mu/\mu^{(1)}$ was discretised into $100$ representative points, then for each value of $\mu/\mu^{(1)}$, $100$ random values of $a$ were numerically generated from the specified Gamma distribution and compared with the inequalities defining the two intervals for values of $\mu/\mu^{(1)}$.  

For the deterministic solution based on the mean value of the nematic parameter, $\underline{a}=\rho_{1}\rho_{2}=2$, the critical value of $\mu/\mu^{(1)}=1.3512$ strictly separates the regions where striping inhomogeneity occurs or not. For the stochastic version, to increase the probability of homogeneous deformations ($P_{2}\approx 1$), one must decrease the values of $\mu/\mu^{(1)}$, whereas shear striping is certain ($P_{1}\approx 1$) only if the model reduces to the neoclassical one (i.e., when $\mu^{(1)}=0$). When $\mu^{(1)}>0$, there will always be competition between the homogeneous and striping deformations. 

To illustrate the effect of the probabilistic parameter $a$, in Figure~\ref{LCE:fig:astoch}, we represent the stochastic shear parameter $\varepsilon_{0}$ and director angle $\theta_{0}$, given by \eqref{LCE:eq:eps0}-\eqref{LCE:eq:theta0}, when $a$ is drawn from a Gamma probability distribution with shape and scale parameters $\rho_{1}=200$ and $\rho_{2}=0.01$, respectively, while $\mu=\mu^{(1)}+\mu^{(2)}=4.05$ is kept deterministic, with $\mu^{(1)}=\mu^{(2)}=4.05/2$. To compare directly the stochastic results with the deterministic ones, we sampled from a distribution where the nematic parameter was set to have the mean value corresponding to the deterministic system. Due to the nonlinear form in which these parameter appear in the model, there will be a significant difference between the mean value solutions and the deterministic solutions. In the particular case when $\mu^{(1)}=0$, this difference was accounted for explicitly by the model proposed in \cite{Verwey:1995:VW,Verwey:1996:VWT}, where a small term was added to the one-term neoclassical form to obtain a mean value expression for the strain-energy function.

\subsection{Simple shear deformation}

We consider only briefly the simple shear deformation of a liquid crystal elastomer where the nematic director $\textbf{n}$ and its reference orientation $\textbf{n}_{0}$ are given by \eqref{LCE:eq:n0ntheta}, and the gradient tensor is equal to \cite{DeSimone:2009:dST} 
\begin{equation}\label{LCE:F:shear}
\textbf{F}=
\left[
\begin{array}{ccc}
a^{-1/6} & 0 & 0\\
0 & a^{-1/6} & \varepsilon\\
0 & 0 & a^{1/3}
\end{array}
\right],
\end{equation}
with $a>1$ and $\varepsilon>0$. In this case,
\begin{equation}\label{LCE:FG:shear}
\textbf{F}\textbf{G}_{0}^{-1}=
\left[
\begin{array}{ccc}
1 & 0 & 0\\
0 & 1 & \varepsilon\\
0 & 0 & 1
\end{array}
\right].
\end{equation}
Noting that \eqref{LCE:F:shear} is obtained by setting $\lambda=a^{-1/6}$ in \eqref{LCE:F:stretch}, when the material is characterised by the MR-type relation \eqref{LCE:eq:W:Fn:MR}, there exists an equilibrium state at $\varepsilon=0$ and $\theta=0$. The associated function $w(\lambda,\varepsilon,\theta)$ defined by \eqref{LCE:eq:W:Fn:MR:stretch} then takes the approximate form \eqref{LCE:eq:W:2order:stretch} with $\lambda=a^{-1/6}$. It follows that the equilibrium state is stable if $\mu^{(2)}=\mu^{(2)}_{1}+\mu^{(2)}_{2}\geq0$ and $\mu^{(1)}=\mu^{(1)}_{1}+\mu^{(1)}_{2}>0$, and neutrally stable if $\mu^{(1)}=0$. Therefore, if $\mu^{(1)}$ and $\mu^{(2)}$ are random variables, the equilibrium solution is stable (or neutrally stable) with 100\% certainty. In this case, the deformation is always homogeneous, and the only difference in the stochastic case compared to the deterministic one consists in the fact that the solution will 'fluctuate' around the mean value. 
 
\section{Conclusion}\label{LCE:sec:conclude}

We studied here shear striping in stretched monodomain nematic elastomers described by constitutive models combining purely elastic and neoclassical-type strain-energy densities. In its general form, the elastic-nematic model can be expressed in terms of invariants, and we considered in particular models with a superposition of an elastic Mooney-Rivlin strain energy on a Mooney-Rivlin-type neoclassical energy, and similarly, a combined Gent-Gent-type strain-energy function. We found that, unlike in the purely neoclassical case, where the inhomogeneous deformation occurs within a constant interval of the extension ratio, independently of the model coefficients, for the elastic-nematic models, the critical interval depends on the ratio between the coefficients of the elastic and the neoclassical part. When the elastic contribution is small, this interval converges to that of the purely neoclassical case, while if the neoclassical contribution is small, then the inhomogeneity interval tends to zero. 

To capture the inherent variability in the physical responses of liquid crystal elastomers, we then extended these models to stochastic-elastic-nematic forms where either the shear modulus $\mu$ or the nematic stretch parameter $a$ are random variables described by non-Gaussian probability density functions at a macroscopic scale. In many applications, the random variation observed in the available data is described by a normal (Gaussian) probability distribution. However, in general, material measurements show a distribution where mean values are low, variances are large, and values cannot be negative. These are important features of the experimental data that cannot be ignored by mathematical models, even though, for small variability, both the normal and Gamma distributions agree well with data collected from material tests \cite{Fitt:2019:FWWM,Mihai:2019a:MDWG}. This guides us to choose the distribution that is suitable also for cases exhibiting large variability, i.e., the Gamma distribution. 

In contrast to the deterministic versions where single-valued critical constants strictly separate homogeneous stretching from inhomogeneous shear striping, for the stochastic models, we obtained a probabilistic interval where the homogeneous and inhomogeneous states compete, in the sense that both have a quantifiable chance to be observed with a given probability. Outside of this critical interval, the deformation is always homogeneous, and it makes little difference whether the model parameters are stochastic or not, although, in the stochastic case, the solution will appear somewhat `dispersed' around the mean value. However, within the probabilistic interval, the variance in the stochastic inhomogeneous solution changes non-uniformly about the mean value. This suggests that, when the variance increases, the average value becomes less significant from the physical point of view. The theoretical approach developed here can be extended to treat other problems, either analytically or computationally, within the same stochastic-nematic framework. 

\appendix
\section{Simple shear as biaxial stretch}\label{LCE:sec:append}

It is instructive to briefly review a classic result of finite strain analysis showing that the simple shear deformation of hyperelastic materials is equivalent to a biaxial stretch (``pure shear'' \cite{Rivlin:1951:VII:RS}) in the principal directions \cite{Ogden:1997}. This result is applicable to nematic elastomer to explain the formation and disappearance of shear stripes analysed in Section~\ref{LCE:sec:striping}.

For a unit cube sample of material subject to the following simple shear deformation,
\begin{equation}\label{LCE:eq:ss}
x_{1}=X_{1}+KX_{2},\qquad x_{2}=X_{2},\qquad x_{3}=X_{3},
\end{equation}
where $K>0$ is the shear constant, the principal stretch ratios $\lambda_{i}=\partial x_{i}/\partial X_{i}$, $i=1,2,3$,  satisfy
\begin{equation}\label{LCE:eq:lambda123}
\lambda_{1}^{2}=1+\frac{K^{2}+K\sqrt{K^{2}+4}}{2}=\lambda^2,\qquad \lambda_{2}^{2}=1+\frac{K^{2}-K\sqrt{K^{2}+4}}{2}=\lambda^{-2},\qquad \lambda_{3}^{2}=1.
\end{equation}
Hence, the simple shear deformation given by \eqref{LCE:eq:ss} is equivalent to a biaxial stretch in the principal directions, with the stretch ratio in the first direction equal to
\begin{equation}\label{LCE:eq:lambda}
\lambda=\frac{K+\sqrt{K^{2}+4}}{2}.
\end{equation}

\begin{figure}[htbp]
	\begin{center}
		\includegraphics[width=0.75\textwidth]{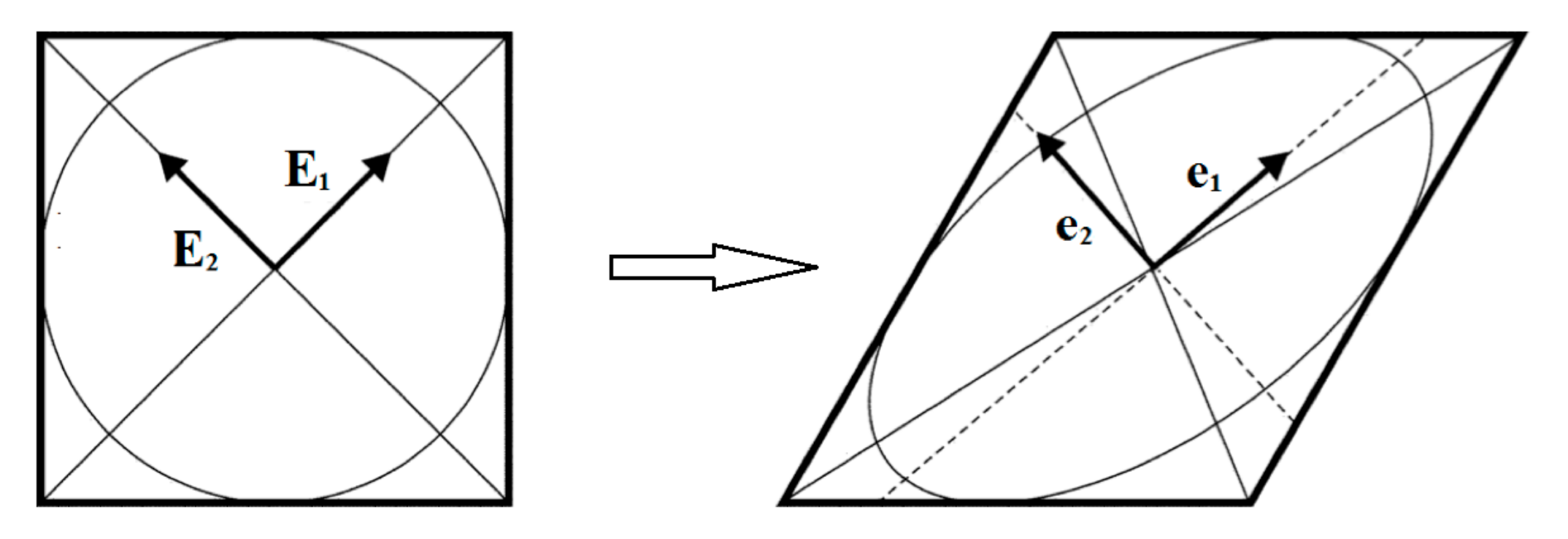}
		\caption{Simple shear of a square section of a material, showing the principal directions before (left) and after (right) deformation.}\label{LCE:fig:draw-ss}
	\end{center}
\end{figure}

Assuming that the following relations between the principal directions $\{\textbf{E}_{i}\}_{i=1,2,3}$ and  $\{\textbf{e}_{i}\}_{i=1,2,3}$, of the reference and current configuration, respectively, hold
\begin{equation}\label{LCE:eq:v123}
\textbf{e}_{1}=\textbf{E}_{1}\cos\alpha+\textbf{E}_{2}\sin\alpha,\qquad
\textbf{e}_{2}=-\textbf{E}_{1}\sin\alpha+\textbf{E}_{2}\cos\alpha, \qquad
\textbf{e}_{3}=\textbf{E}_{3},
\end{equation}
with the angle $\alpha$ satisfying
\begin{equation}\label{LCE:eq:alpha}
\tan\alpha=\frac{2}{K+\sqrt{K^2+4}},\qquad 0<\alpha<\frac{\pi}{4},
\end{equation}
by the polar decomposition theorem, it follows that
\begin{equation}\label{LCE:eq:PD}
\textbf{B}=\textbf{F}\textbf{F}^{T}=\textbf{R}\textbf{U}^2\textbf{R}^{T},
\end{equation}
where
\[
\textbf{F}=\left[
\begin{array}{ccc}
1 & K & 0\\
0 & 1 & 0\\
0 & 0 & 1
\end{array}
\right],
\qquad
\textbf{R}=\left[
\begin{array}{ccc}
\cos\alpha & -\sin\alpha & 0\\
\sin\alpha & \cos\alpha & 0\\
0 & 0 & 1
\end{array}
\right],
\qquad
\textbf{U}=\left[
\begin{array}{ccc}
\lambda & 0 & 0\\
0 & \lambda^{-1} & 0\\
0 & 0 & 1
\end{array}
\right].
\]
The geometric interpretation of the simple shear deformation \eqref{LCE:eq:ss} is that, for any fixed $0<X_{3}<1$, the circle inscribed in the unit square  $[0,1]\times[0,1]$ deforms into an ellipse with the major axis along the principal direction $\textbf{e}_{1}$ and eccentricity $\lambda^2$. Specifically, as
\[
\frac{1}{K+1}<\tan\alpha<1\qquad \mbox{for}\quad 0<K<\infty,
\]
the principal axes are situated between the diagonals of the undeformed and sheared square, respectively (see Figure~\ref{LCE:fig:draw-ss}). In the limiting case when $K\to0$, since $\tan\alpha\approx1$, the Eulerian directions  are along the diagonals of the unit square in the undeformed state, and coincide with the Lagrangian directions, while if $K\to\infty$, then $\tan\alpha\approx 1/(K+1)$, i.e., the first principal direction is effectively on the longer diagonal of the deformed square.

\begin{figure}[htbp]
	\begin{center}
		\includegraphics[width=0.75\textwidth]{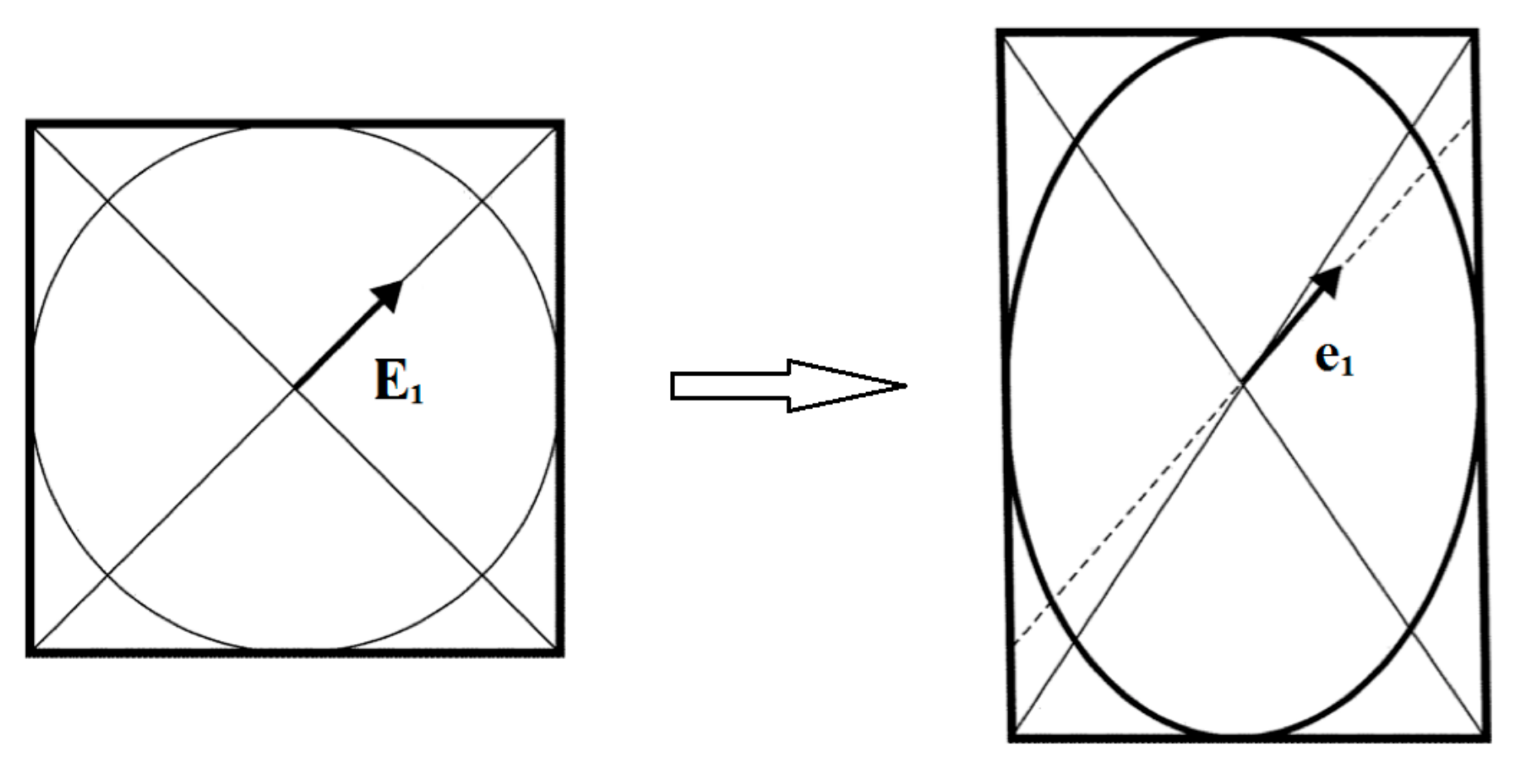}
		\caption{Biaxial stretch of a square section of a material, showing the first principal direction of the associated simple shear, before (left) and after (right) deformation.}\label{LCE:fig:draw-bs}
	\end{center}
\end{figure}

Conversely, by the decomposition (\ref{LCE:eq:PD}), any biaxial stretch of the form
\begin{equation}\label{LCE:eq:bs}
x_{1}=\lambda X_{1},\qquad x_{2}=\lambda^{-1}X_{2},\qquad x_{3}=X_{3},
\end{equation}
where $\lambda>1$, is equivalent to a simple shear with amount of shear $K=\lambda-\lambda^{-1}$ at a relative angle $\alpha$, such that $\tan\alpha=\lambda^{-1}$ (see Figure~\ref{LCE:fig:draw-bs}). In the linear elastic limit, $\lambda\to 1$, the biaxial stretch (\ref{LCE:eq:bs}) is equivalent to a simple shear at a relative angle $\alpha\approx\pi/4$.

For the liquid crystal elastomers presenting shear stripes during biaxial stretch, the striping pattern develops due to the gradual rotation of the nematic director from the initial perpendicular direction to the final parallel direction to elongation. In view of the above derivation, for a shear stripe of length $1$ and width $\delta\in(0,1)$, the first principal direction will be `close' to the longer diagonal of a sheared square with the side equal to the width of the stripe. As macroscopic stretching progresses, so does the equivalent shear deformation. Locally, by \eqref{LCE:eq:alpha}, when the nematic director becomes (almost) parallel to the longitudinal direction of the stripe, the corresponding simple shear has an infinitely large parameter, $K\to\infty$. Simultaneously, in the macroscopic deformation, the effective stretch ratio is $\lambda$ approaches the upper bound given in \eqref{LCE:eq:bounds2}, hence, the associated shear parameter, $k=\lambda-\lambda^{-1}$, is finite and relatively small. From the relation $k=K\delta$, we infer that, as the nematic director rotates, the width of each stripe decreases to zero ($\delta=k/K\to 0$), and eventually the material deforms homogeneously.

\paragraph{Acknowledgement.} The support by the Engineering and Physical Sciences Research Council of Great Britain under research grants EP/R020205/1 to Alain Goriely and EP/S028870/1 to L. Angela Mihai is gratefully acknowledged.


\end{document}